\documentclass[twocolumn,aps,prb,tightenlines,amsfonts,notitlepage]{revtex4-1}
\usepackage[font=footnotesize,labelfont=sf,justification=raggedright]{caption}
\usepackage{graphicx,amsmath,color,amssymb,bm}
\usepackage[margin=1.5cm]{geometry}

\usepackage[colorlinks=true,linkcolor=blue,citecolor=red]{hyperref}

\newcommand{\smhead}[1]{\hspace{0.1cm}\newline\noindent\textbf{#1}\newline}
\begin{document}
\title{Automated calculation and convergence of defect transport tensors}
\author{Thomas D Swinburne}
\email{swinburne@cinam.univ-mrs.fr}
\affiliation{Aix-Marseille Universit\'{e}, CNRS, CINaM UMR 7325, Campus de Luminy, 13288 Marseille, France}
\author{Danny Perez}
\affiliation{Theoretical Division T-1, Los Alamos National Laboratory, Los Alamos, NM, 87545, USA}
\date{\today}
\begin{abstract}
	Defect transport is a key process in materials science and catalysis, but as migration mechanisms
	are often too complex to enumerate \textit{a priori}, calculation of transport tensors typically have no measure of
	convergence and require significant end user intervention. These two bottlenecks prevent high-throughput
	implementations essential to propagate model-form uncertainty from interatomic interactions to
	predictive simulations. In order to address these issues, we extend a massively parallel accelerated sampling scheme,
	autonomously controlled by Bayesian estimators of statewise sampling completeness, to build atomistic kinetic
	Monte Carlo models on a state space irreducible under exchange and space group symmetries.
	Focusing on isolated defects, we derive analytic expressions for defect transport tensors
	and provide a convergence metric by calculating the Kullback-Leiber divergence across the
	ensemble of diffusion processes consistent with the sampling uncertainty.
	The autonomy and efficacy of the method is demonstrated on surface trimers in tungsten and hexa-interstitials in magnesium oxide,
	both of which exhibit complex, correlated migration mechanisms.
\end{abstract}
\maketitle
\smhead{Introduction}
	The migration and transformation of intrinsic and extrinsic complex crystal defects plays a central role in numerous materials science and chemistry
	phenomena such as post-irradiation annealing\cite{fu2005}, plasma surface interactions\cite{sefta2013}, active site formation for heterogeneous
	catalysis\cite{norskov2009,boller2019} or mechanical properties of concentrated solid solution alloys\cite{osetsky2018}.

	The atomistic mechanisms available to nanoscale defects are highly heterogeneous with defect size and impossible to divine \textit{a priori} due to the routine presence
	of complex multi-atom transformations\cite{sorensen2000,uberuaga2005,uberuaga2007,perez2009,beland2011}.
	Whilst these effects typically becomes less volatile with increasing defect size, phenomenological or higher-scale models can only capture the defect dynamics
	if suitable parameters can be calculated\cite{dezerald2015first,swinburne2016,alexander2016}.
	In addition, nanoscale clusters are typically the most mobile and thus often have a much greater influence on macroscopic phenomena than slower-moving larger defects.
	In the general case, atomistic mechanisms must be discovered through unbiased dynamic\cite{voter1997,voter1998parallel,TAD,perez2009,chatterjee2015uncertainty,chill2014}
	or static\cite{dimer,beland2011,wales2002discrete} sampling approaches. When the true dynamics can be characterized as rare transitions between metastable basins
	on the energy landscape\cite{wales_energy_2003},
	the basin-to-basin dynamics can be mapped to a continuous time Markov chain \cite{lebris2012,Lelièvre2018}, which forms the theoretical foundation of atomistic
	kinetic Monte Carlo (akMC) methods\cite{henkelman2017}, of which on-lattice kMC is a subclass. The resulting model can then be stochastically or in some cases
	(such as that presented here) analytically integrated to extract observables of interest.

	A well recognized problem is that an akMC model will in general have an incomplete catalogue of available mechanisms due to finite amount of sampling, and this
	can produce catastrophically erroneous predictions if important mechanisms are omitted\cite{chill2014,chatterjee2015uncertainty,aristoff2016,swinburne2018b}.
	Sampling adequacy is often assesed \textit{qualitatively} using the domain expertise of simulation practicioners. Whilst this approach has undoubtably yeilded
	significant successes, it requires significant end user analysis for each system under study. This has a punitive impact on the feasibility of automating complex
	materials simulations on massively parallel computational resources, where the required decision frequency rapidly exceeds
	practical human limits. Further, in absence of \textit{quantitative} uncertainty quantification approaches, assessing the reliability of
	meso or macro-scale predictions is extremely challenging.

	For example, the system sizes required even for small defect clusters render \textit{ab initio} calculation unfeasible.
	As a result, interatomic potential models must be used which induces an additional model form uncertainty.
	The development of interatomic potentials has been revolutionized in recent
	years through the use of linear-in-descriptor or neural network regression techniques
	\cite{bartok2017,cosmin2019,shapeev2016moment}. These approaches offer
	a natural encoding of model form uncertainty through isosurfaces of the cost function used
	for potential parametrization. High-throughput calculations
	are essential in this context to enable the systematic propagation of this uncertainty on
	interatomic interactions to the observables of scientific interest, without prohibitive hours of end-user analysis.

	In this contribution we present an autonomous, highly scalable sampling scheme to efficiently calculate defect
	diffusion tensors with quantified uncertainty on the sampling completeness.
	We demonstrate the ability to discover complex migration behaviours of defects in tungsten and magnesium oxide,
	and show how the quantified uncertainty can be used to rapidly yeild well defined convergence measures.
	Our approach enables high-throughput workflows to rapidly discover, converge and analyze complex kinetic properties of
	defect structures in surrogate energy landscapes, with minimal end user involvement.

\smhead{Results}\label{sec:TAMMBER}
	In previous work\cite{swinburne2018b} we introduced \texttt{TAMMBER}, a massively parallel accelerated sampling scheme
	whose formal objective is to exhaustively sample the set of all metastable minima $\mathcal{M}$ and all interminima
	transitions $\mathcal{T}$ for a given system. \texttt{TAMMBER} builds a matrix ${\rm K}_{ji}$ of $j\leftarrow i$ rates
	in the known state space and a vector ${\rm k}_i^u$ which estimates the `unknown' (as yet undiscovered) escape rate
	from each state $i$, encoded as transitions
	to an absorbing sink\cite{boulougouris2005monte,chill2014,bhoutekar2017new,chatterjee2015uncertainty,swinburne2018b}
	(methods). States are identified by constructing a connectivity
	graph from a minimized configuration, which is then hashed to produce a pseudounique integer
	label for each configuration\cite{beland2011,perez2015}.
	The presence of nonzero ${\rm k}^u$ causes trajectories to leave the known state space, giving a natual measure of model validity,
	the expected residence time\cite{swinburne2018b} in the known state space
	\begin{equation}
	\langle\tau_{res}\rangle = {\bf 1}\left[{\bf K}^{tot}-{\bf K}\right]^{-1}{\bf P}_0,\label{tau}
	\end{equation}
	where ${\mathrm K}_{ji}^{tot} = \delta_{ji}\left[{\rm k}_i^u+\sum_l{\rm K}_{li}\right]$, ${\bf P}_0$ is the initial
	distribution and ${\bf 1}$ is a row vector of ones.
	In practice, truly comprehensive sampling of $\mathcal{M}$ and $\mathcal{T}$ is often impractical as their size grows
	exponentially with the number of atoms in the system. However, $\mathcal{M}$ and $\mathcal{T}$ are both highly
	reducible under exchanges of indistinguishable atoms $\mathcal{R}$, lattice vector
	translations $\mathcal{S}$ and space group symmetries $\mathcal{G}$, especially in the case of quasi-zero-dimensional
	defects.
	It is known that configurations which are degenerate (in the periodic minimum image
	sense) under these operations will have isomorphic connectivity graphs\cite{beland2011,perez2015}, with isomorphisms
	that can be efficiently	calculated\cite{NAUTY}. We exploit this reducibility to make a partition of $\mathcal{M,T}$
	into equally sized subsets
	\begin{equation}
		\mathcal{M}=\bigcup_{p\in[1,P]}\mathcal{M}_p,
		\quad\quad
		\mathcal{T}=\bigcup_{p,q\in[1,P]^2}\mathcal{T}_{pq},\label{sgp}
	\end{equation}
	where every member in a partition $\mathcal{M}_p$ is identical to all other members within a reindexing in
	$\mathcal{R}$ and translation by ${\bf t}\in\mathcal{S}$. Similarly, the subset of transition
	$\mathcal{T}_{p_0q}\subset\mathcal{T}_{pq}$ that take a configuration $p_0\in\mathcal{M}_p$ to
	members in $\mathcal{M}_q$ is identical to any other subset $\mathcal{T}_{p_iq}$ under application of $\mathcal{R}$ and
	$\mathcal{S}$.
	To exploit space group symmetries, we first note that every defect structure has a symmetry group
	$\mathcal{G}_p\subseteq\mathcal{G}$, a subgroup of $\mathcal{G}$ containing	at least the identity	operation, whose
	order $\#\mathcal{G}_p$ (number of elements) factorizes the order $\#\mathcal{G}$ of $\mathcal{G}$.
	When acting on every member of $\mathcal{M}_p$ with some space group symmetry
	${\bf G}\in\mathcal{G}$ to produce a new set of configurations ${\bf G}\mathcal{M}_p$, there are two possible outcomes:
	${\bf G}\mathcal{M}_p=\mathcal{M}_p$ if ${\bf G}\in\mathcal{G}_p$ or ${\bf G}\mathcal{M}_p=\mathcal{M}_q$
	(for some $q\neq p$) if ${\bf G}\not\in\mathcal{G}_p$.
	In the latter case we have acheived a further reduction in the
	number of partitions we need to consider, as $\mathcal{M}_q$ can be generated from $\mathcal{M}_p$ through the action
	of $\bf G$.
	However, this only defined up to postmultiplication by ${\bf G}_p\in\mathcal{G}_p$, as
	${\bf G}{\bf G}_p\mathcal{M}_p={\bf G}\mathcal{M}_p$. To resolve this ambiguity we generate all
	possible (left) cosets\cite{scott2012group} with respect to $\mathcal{G}_p$, defined for some ${\bf G}\in\mathcal{G}$ as
	$\{{\bf G}{\bf G}_p : {\bf G}_p\in\mathcal{G}_p\}$, i.e. the set of operations formed by postmultplying ${\bf G}$
	by every element in $\mathcal{G}_p$\footnote{right coset are formed through premultiplication}.
	We can always form a set $\bar{\mathcal{G}}_p$ of exactly $\#\mathcal{G}/\#\mathcal{G}_p$ nonequivalent cosets of size $\#\mathcal{G}_p$,
	where each member of $\mathcal{G}$ appears in exactly one coset\cite{scott2012group}.
	Clearly, the elements of $\mathcal{G}_p$ form one of these cosets, whilst
	acting on $\mathcal{M}_p$ with any element from a	given coset will give the same result.
	We can then index the $\#\bar{\mathcal{G}}_p$ nonequivalent partitions $\{\mathcal{M}_q\}$ which can be generated from $\mathcal{M}_p$ by the
	coset which generates them; drawing some operation ${\bf G}_{qp}$ from each coset we
	can thus make the further compression $\mathcal{M}=\bigcup_{p=1}^{P_G}\bigcup_{q}{\bf G}_{qp}\mathcal{M}_p$, where
	$P_G\leq P$, with analagous partitioning of $\mathcal{T}_{pq}$.

\begin{figure}
		\includegraphics[width=\columnwidth]{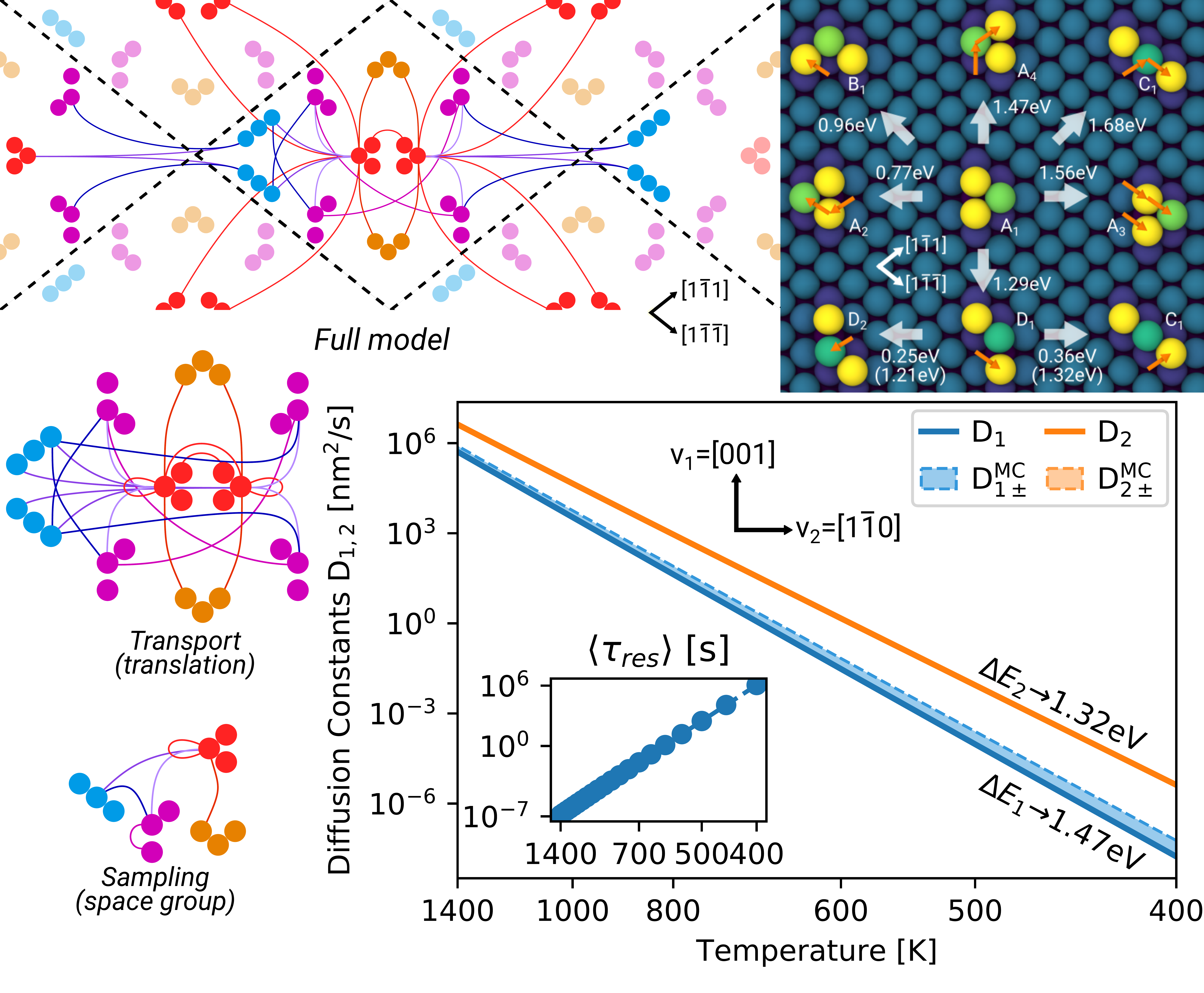}
		\caption{Compressed sampling methodology applied to trimer migration on W$(110)$, which has symmetry $C_{2v}$.
		Top Left: The lowest energy irreducible states and found transition mechanisms. Atoms colored by
		centrosymmetry\cite{OVITO}.
		Top Right: Cartoon of the full akMC state space and those irreducible under space and translation symmetries. Isomorphic states have the same color.
		One primitive unit cell with self transitions (closed loops) is sufficient to build a Markov model for transport,
		whilst the set of states
		irreducible under all space group symmetries is optimal for dicovering new states and transitions through accelerated sampling.
		Bottom:
		Diffusion tensor eigenvalues ${\rm D}_{1}$, ${\rm D}_{2}$, as determined by (\ref{trajD}), over a range of
		temperatures. The Monte Carlo bounds ${\rm D}^{MC}_{1\pm}$,
		${\rm D}^{MC}_{2\pm}$ are described in the main text. Effective Arrhenius slopes at low tempeature are given.
		Inset: $\langle\tau_{res}\rangle$ with temperature.
		}
		\label{Wmodel}
	\end{figure}

	We therefore only need to sample the exit transitions from a maximum of $P_G$ configurations, as all others can be
	generated by the known symmetric relations. By focusing sampling resources on a much smaller number of possible states,
	the unknown rates decrease at a much faster rate, with a subsequent increase in the validity timescales for the
	constructed Markov chain. As in \texttt{TAMMBER} the unknown rates at low temperatures decrease as an inverse power of
	the MD time invested at high temperatures, state-space compression can yield extremely large benefits in practice. This
	maximally compressed representation is thus used for sampling, which is then partially decompressed into the
	partitioning (\ref{sgp}) of $P$ distinct states irreducible under translation for the calculation of transport tensors,
	as illustrated in figure (\ref{Wmodel}). Whilst connectivity graph isomorphisms have been used to identitify
	already seen defect structures\cite{beland2011}, the production of Markov chains with complete or partial
	irreducibility for sampling or model construction is novel to the best of our knowledge.
	As an example of the efficiencies this approach affords, in previous work we investigated the breakup of two
	interstitial defects in bcc iron\cite{swinburne2018b}, obtaining a Markov model on the uncompressed (reducible) state
	space with $\langle\tau_{res}\rangle\sim80s$ at 300K, an insufficent duration to make confident predictions on the
	breakup mechanism. Despite the relatively low symmetry of this defect system, we show in the supplementary material
	that the compressed sampling scheme described above yields a Markov model with
	$\langle\tau_{res}\rangle\sim5\times10^6s$ with only 75\% of the computational effort, allowing convergence in the
	model predictions. The isolated point defects we consider in this work are typically of even higher symmetry, giving
	correspondingly greater efficiencies.
\smhead{Evaluation of transport tensors}
	To define the drift and diffusion tensors, we first require a defect position in the supercell for each configuration.
	To remove ambiguities from periodic boundary conditions we carve out a defective region of some configuration in each
	partition by thresholding some structural descriptor, here the centrosymmetry, then take a descriptor-weighted center of mass.
	As we allocate sampling to increase the residence time, the threshold value can be freely determined in post-processing;
	typically a single value is suitable for the same material system. It is also possible to determine the defect position through
	analysis of a sufficiently large set of mutally isomorphic configurations, but we find the descriptor-based approach to be simple, efficient and reliable.

	A periodic minimum image displacement vector ${\bf d}_{lm}$ for any $l\in\mathcal{M}_p\to m\in\mathcal{M}_q$ transition can then be found
	through application of the known rigid transformation from the sampled configurations.
	Importantly, transitions between states within the same space group partition are then represented as \textit{self transitions}
	$l\in\mathcal{M}_p\to m\in\mathcal{M}_p$ (closed loops in figure \ref{Wmodel}).

	We can then construct an irreducible group of states from the space group partitioning of
	equation (\ref{sgp}), with one state indexed by $p\in[1,P]$ for each set $\mathcal{M}_p$.
	All the transitions from a state $p$ to states $q$ in the uncompressed model are mapped to their compressed states,
	forming a (possibly empty) set $\mathcal{C}_{qp}=\{{\rm k}_l,{\bf d}_l\}$ of transitions and displacement vectors.
	The compressed rate matrix is then defined as ${\rm K}^{c}_{qp} = \sum_{l\in\mathcal{C}_{qp}} {\rm k}_{l}$, with a statewise
	total escape rate matrix ${\rm K}^{tot,c}_{qp} = \delta_{qp}({\rm k}^{u,c}_p+\sum_r{\rm K}^c_{rp})$,
	where the superscript $c$ indicates we consider rates in the compressed model. The
	unknown rates ${\rm k}^{u,c}_{p}$ are taken from the maximally compressed representation,
	with the same unknown rate for symmetrically equivalent partitions.
	One could then generate akMC trajectories\cite{bortz1975} with a matrix and vector of branching probabilities
	${\bf B}^c \equiv {\bf K}^c[{\bf K}^{tot,c}]^{-1}$ and ${\bf B}^{u,c} \equiv {\bf k}^{u,c}[{\bf K}^{tot,c}]^{-1}$,
	accumulating a total displacement ${\bf x}=\sum {\bf d}_l$ in a residence time $\tau_{res} = \sum \delta\tau_l$ before
	absorbtion. In the supplementary material we show that the drift and diffusion coefficients ${\bm\mu},{\bf D}$ can then
	be extracted from the relations $\langle{\bf x}\rangle=\langle\tau_{res}\rangle{\bm\mu}$ and
	$\langle{\bf x}\otimes{\bf x}\rangle=2\langle\tau_{res}\rangle{\bf D}+\langle\tau^2_{res}\rangle{\bm\mu}\otimes{\bm\mu}$.
	However, it is also possible to analytically evaluate averages over all possible pathways using a
	`displacement generating function'
	\begin{equation}
		Z({\bm\lambda},{\bf P}_0)=
		{\bf B}^{u,c}
		{\bf G}^c({\bm\lambda})
		%\left[\mathbb{I}-{\bf B}^c({\bm\lambda})\right]^{-1}
		{\bf P}_0
		%\hat{\bm\pi}^{QSD}%
		,\quad
		[{\bf B}^c({\bm\lambda})]_{qp} \equiv
		\sum_{l\in\mathcal{C}_{qp}} \frac{{\rm k}_{l}e^{-{\bm\lambda}\cdot{\bf d}_{l}}}{{\rm K}^{tot,c}_{pp}},
		\label{Z}
	\end{equation}
	where ${\bf G}^c({\bm\lambda})=\sum^\infty_{n=0}\left[{\bf B}^c({\bm\lambda})\right]^n=\left[\mathbb{I}-{\bf B}^c({\bm\lambda})\right]^{-1}$.
	Moments of the total displacement can then be written\\
	$\langle{\bf x}\rangle=-\partial_{\bm\lambda}Z({\bm\lambda,{\bf P}_0})|_{\bm\lambda=\bf0}$
	and	$\langle{\bf x}\otimes{\bf x}\rangle=\partial_{\bm\lambda}\otimes\partial_{\bm\lambda}Z({\bm\lambda,{\bf P}_0})|_{\bm\lambda=\bf0}$ (supplementary material).
	Whilst (\ref{Z}) could be used for any choice of initial condition, we note that in the well sampled limit ${\rm k}_p^{u,c}\ll\sum_{r}{\rm K}^c_{rp}$,
	which is necessary but not sufficient for global convergence, the matrix $\left[{\bf K}^{tot,c}-{\bf K}^c\right]$ will have a spectral gap,
	with one eigenvalue $0<\nu_0\ll\nu_1<\nu_2...$ much smaller than all others\cite{lebris2012}.
	The right eigenvector for $\nu_0$ is the \textit{quasistationary} distribution (QSD) ${\bm\pi}^{QSD}$ in the known state space\cite{lebris2012}, the limiting
	distribution conditional on not absorbing for an arbitrary long time, which as ${\bf k}^{u,c}\to{\bf0}$ becomes the Boltzmann distribution, ${\bm\pi}^{QSD}\to\hat{\bm\pi}$.
	%{If the system exhbits broken ergodicity, typically due to incomplete sampling, the QSD will be degenerate, a point we return to below.}
	As the QSD is the longest-lived mode and transport coefficients are defined as the limit of infinitely long trajectories, it is natural to set
	${\bf P}_0=\hat{\bm\pi}^{QSD}={\bm\pi}^{QSD}/({\bf1}{\bm\pi}^{QSD})$ to eliminate the influence of initial conditions.
	With this choice, it is simple to show that $\langle\tau^n_{res}\rangle=n!/\nu^n_0$ and the
	expected drift and diffusion coefficients emerge as (supplementary material)
	\begin{align}
		{\bm\mu}({\bf k}^{u,c}) &\equiv  \langle\tau_{res}\rangle^{-1}\partial_{\bm\lambda}\left[1/Z({\bm\lambda},\hat{\bm\pi}^{QSD})\right]
		\Big|_{{\bm\lambda}={\bf0}},%,{\bf P}_0=\hat{\bm\pi}^{QSD}},
		\label{traj}
		\\
		{\bf D}({\bf k}^{u,c}) &\equiv
		\frac{1}{2}
		\langle\tau_{res}\rangle^{-1}
		\partial_{\bm\lambda}\otimes\partial_{\bm\lambda}\left[1/Z({\bm\lambda},\hat{\bm\pi}^{QSD})\right]
		\Big|_{{\bm\lambda}={\bf0}}.%,{\bf P}_0=\hat{\bm\pi}^{QSD}}.
		\label{trajD}
	\end{align}
	Equations (\ref{traj},\ref{trajD}) are a central result of this contribution, expressions for the drift and diffusion tensors of an
	arbitrarily complex diffusion process in a periodic system, autonomously constructed in a massively parallel sampling scheme,
	which crucially are dependent on `unknown' rates ${\bf k}^{u,c}$ that robustly quantify sampling incompleteness.
	In the supplementary material we show the limiting expressions ${\bm\mu}\equiv\lim_{{\bf k}^{u,c}\to{\bf0}}{\bm\mu}({\bf k}^{u,c})$
	and ${\bf D}\equiv\lim_{{\bf k}^{u,c}\to{\bf0}}{\bf D}({\bf k}^{u,c})$ reduce to expressions obtained in previous derivations
	on multistate diffusion in periodic media\cite{Trinkle2018,landman1979stochastic}, with all uncorrelated and correlated
	contributions that are essential to capture complex diffusion pathways. In the remainder we focus on systems obeying detailed balance,
	where ${\bm\mu}={\bf0}$	and the correlated contribution to the diffusivity is always nonpositive.

\smhead{Convergence of the diffusivity}
	The central novelty of (\ref{trajD}) is that ${\bf D}({\bf k}^{u,c})$ estimates the diffusion tensor over all
	possible trajectories in the known state space before exit due to the unknown rates. Possible changes to the diffusivity
	under the discovery of additional transition rates between \textit{known} states can then be bounded; the rate matrix
	${\bf K}^c$ can be modified by an additional rates matrix $\delta{\bf K}^c$ which must satisfy detailed balance and
	not increase the total exit rate from	each known state $p$ by more than the unknown rate ${\rm k}^{u,c}_p$, meaning
	$\left[{\bf 1}\delta{\bf K}^c\right]_p\leq{\rm k}^{u,c}_p$. We have desiged a Monte Carlo procedure to sample the
	space of permissible $\delta{\bf K}^c$ (see supplementary material) which typically requires less than a core-minute
	for the systems studied here and is trivally parallelizable.
	We discuss the sensitivity to the discovery of additional states at the end of this section.

	To analyze the ensemble of diffusion tensors produced in the Monte Carlo procedure, we diagonalize the $3\times 3$ diffusion matrix, producing eigenvalues $\{D_l\},l\in[1,3]$.
	The Monte Carlo procedure yeilds upper and lower bounds ${\rm D}_{l\pm}$ for each of the eigenvalues from which a convergence metric can
	be obtained; in practice, we find these bounds to be highly asymmetric, with ${\rm D}_{l-}\to{\rm D}_l$ when we constrain the Monte Carlo procedure to
	consider only the addition of new transition rates without any additional state. In addition, the eigenbasis undergoes negligible changes close to
	convergence, meaning that to a high degree of accuracy the matrices ${\bf D}_\pm$ (with eigenvalues ${\rm D}_{l\pm}$) can be simultaneously diagonalized.
	To produce a dimensionless convergence measure, consider the fundamental solutions $\rho({\bf x},{t})$, $\rho_\pm({\bf x},{t})$ to the three
	dimensional diffusion equation with a diffusion tensor ${\bf D}$, ${\bf D}_\pm$. A natural measure is the Kullback-Leibler divergence\cite{cover2012}
	$\mathcal{R}_\pm \equiv \int{\rm d}{\bf x}\rho\ln\rho/\rho_\pm$, which for diffusion is time independent, reading
	$\mathcal{R}_\pm=\mathrm{ Tr}\left({\bf D}^{-1}{\bf D}_\pm\right)/2-3/2+\ln\sqrt{\left\lVert{\bf D}^{-1}{\bf D}_\pm\right\rVert}$.
	Our convergence measure is the spread $\delta\mathcal{R}\equiv\mathcal{R}_+-\mathcal{R}_-\geq0$, which when ${\bf D}_\pm$ and ${\bf D}$ can be
	simultaneously diagonalized reads
	\begin{equation}
		\delta\mathcal{R}=\sum_{l}
		\frac{{\rm D}_{l+}-{\rm D}_{l-}}{2{\rm D}_{l}} +
		\ln\sqrt{\frac{{\rm D}_{l+}}{{\rm D}_{l-}}}
		\to
		\sum_{l}
		\frac{{\rm D}_{l+}-{\rm D}_{l-}}{{\rm D}_{l}},
	\end{equation}
	where the limit applies close to convergence, being the leading order expansion in ${\rm D}_{l\pm}-{\rm D}_l$.
	As $\delta\mathcal{R}=0$ for ${\bf D}_+={\bf D}_-$, we have a well defined dimensionless convergence metric ideal for autonomous implementation,
	which has an informative limit, namely the relative spread in eigenvalues consistent with the sampling uncertainty.

	It is clear that the diffusivity cannot be globally bounded against the discovery of some new set of states which are free to possess arbitrary transport properties.
	However, as transition rates to such a set of states are	bounded by the ${\bf k}^u$, the Monte Carlo prodedure outined
	above does characterizes transport behaviour in the known state space over timescales of order
	$\langle\tau_{res}\rangle$. Convergence to the large $\langle \tau_{res}\rangle$ limit can be accelerated by seeding
	the \texttt{TAMMBER} procedure with as many states as possible, which are free to be completely
	disconnected\footnote{In this case we have a separate QSD ${\bm\pi}^{QSD}_s$ for each connected subnetwork $s$, which
	only has support on that subnetwork. We then calculate ${\bf D}({\bf k}^{u,c})$ with ${\bf P}_0=\sum_s{\bm\pi}^{QSD}_s$, and the limiting eigenspectrum is degenerate}.
	The integration of automated structural search algorithms\cite{wales_energy_2003,marinica2012} into the present
	workflow will be the subject of a future work. Of
	course, as \texttt{TAMMBER} generates thousands of high temperature molecular dynamics trajectories across the known
	state space it is an automated global minimum search method,	with $\langle \tau_{res}\rangle$ acting as quantitative
	measure of sampling quality when the system is ergodic. As a result, whilst in common with all theoretical studies on
	high dimensional landscapes we cannot provide bounds on global minima, we can provide a key
	uncertainty quantification on the validity of our findings, namely a rigorous prediction timescale from an arbitrary
	distribution on the known state space with a corresponding bound on transport coefficients. We emphasize that
	the inability to assign bounds on global minima searches applies equally to human guided sampling or the present approach. The convergence metrics we provide are thus a valuable analytical tool which removes ambiguities inherent to traditional methods in addition to enabling a fully automated workflow suitable for high-throughput computation.
\smhead{Application to trimer diffusion on W$(110)$}\label{sec:APP}
	The diffusion of adatom clusters is a fundamental process in surface science, and has recently been conjectured to play a crucial role
	in the formation of complex `fuzzy' surface morphologies during plasma exposure in nuclear fusion reactors\cite{yang2019}.
	We focus here on the trimer defect a demonstrative case study; comprehensive high-throughput investigations for which the present approach is
	designed will be presented elsewhere. In the present case \texttt{TAMMBER} was initialized with state $A_1$ in figure \ref{Wmodel}, then run for 8 hours on
	144 cores using the \texttt{EAM4}	embedded atom method potential by Marinica \textit{et al.}\cite{marinica2013}, covering a temperature range of 400K-1400K. The corresponding values of $\langle\tau_{res}\rangle$ as a function of temperature are also shown in Fig.\ \ref{Wmodel}.

	The resultant diffusion behaviour is highly correlated,	with many $A_1\leftrightarrow A_2$ and $D_1\leftrightarrow D_2$ transitions in particular.
	However, the overall system does not exhibit a clear `superbasin-to-superbasin' diffusion mechanism, meaning access to the full highly correlated trajectory
	ensemble is essential to extract accurate transport coefficients.
	As shown in Fig.\ \ref{Wmodel}, the eigenvalues ${\rm D}_1,{\rm D}_2$ (${\rm D}_3\to0$ for surface diffusion) show a high degree of convergence under the Monte Carlosensitivity procedure, with	$\delta\mathcal{R}\leq2$ at all temperatures. The eigenvectors for ${\rm D}_1$ and ${\rm D}_2$ were found to be ${\bf v}_1=[001]$ and
	${\bf v}_2=[1\bar{1}0]$. To look for dominant pathways, we form a weighted graph from the connectivity of four
	primitive unit cells, where the graph edges are weighted by the corrsponding saddle point
	energy\cite{swinburne2018b,wales_energy_2003}. Dijstra's shortest path algorithm\cite{dijkstra59} was then used to
	identify the dominant pathways. In agreement with the found Arrhenius slopes, migration along ${\bf v}_2$ is dominated
	by ${A}_1\to{D}_1\to{C}_1\to{D}_2\to{A}_5$ paths, with a well defined activation energy of $\Delta E_{1} = 1.32{\rm eV}$ at the lower temperatures. Migation along ${\bf v}_1$ is similarly dominated by $A_1\to{A_4}$ paths at lower temperatures, but $A_1\to{A_3}$ paths have a growing contribution with temperature, giving a weak nonlinearity to the
	Arrhenius gradient $\Delta E(\beta)=-\partial_\beta\ln|D|$. An in depth study of this procedure, and its role in a fully automated workflow, will be the subject of future work.
	We note that a recent study\cite{yang2019} of surface island
	diffusion on $W(110)$ at 1000K using a different interatomic potential\cite{juslin2013} reported trimer migration via
	$A_1\to A_3$, which is accounted for in the present study but is not found to be the dominant mechanism. The
	ability to efficiently resolve such ambiguities, without prohibitive person hours, is a
	key advantage of methodology presented here.

	\begin{figure}
		\includegraphics[width=\columnwidth]{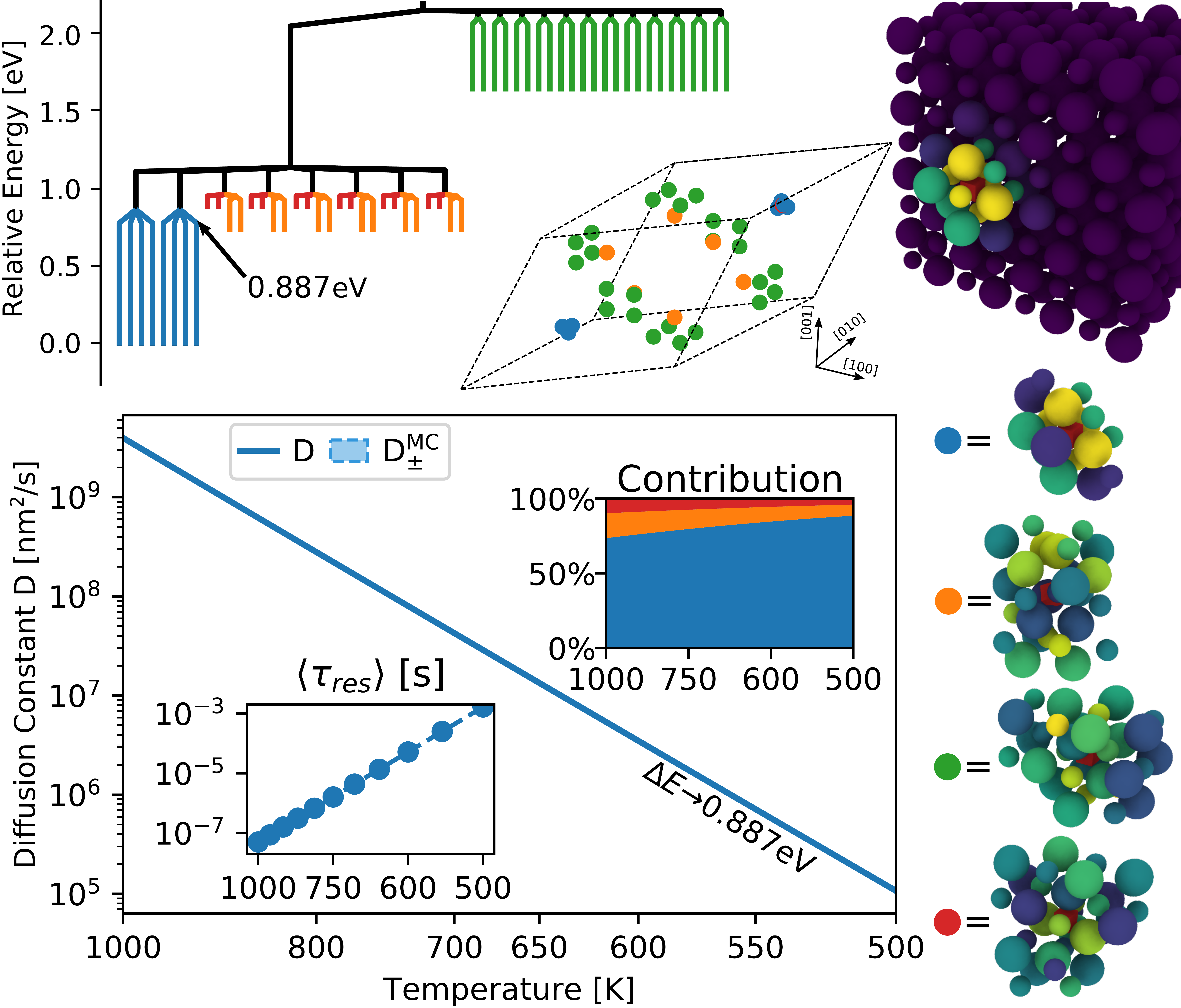}
		\caption{Sampling of hexa-interstitial in MgO. Top left: Disconnectivity graph\cite{wales_energy_2003} between states irreducible under translation. The
		very low diffusion barriers (<0.1eV) observed in previous work\cite{uberuaga2005} can be seen. Top center: Non-equivalent positions in the	primitive unit
		cell for the four irreducible states. Top right: Cross section of the lowest energy state. Atoms colored by centrosymmetry\cite{OVITO}.
		Right: Atomic struture of the four lowest energy irreducible states. Bottom left: Single eigenvalue of the diffusion tensor at various temperatures. Error
		bounds, calculated through the Monte Carlo procedure detailed in the main text, are not visible on the presented scale. Insets show
		$\langle\tau_{res}\rangle$ at various temperatures and the contribution to the overall diffusivity from self transitions within each state. The fast
		diffusing states have increasing contribution at high temperatures as they are occipied more frequently.}
		\label{MgO}
	\end{figure}
\smhead{Application to hexa-interstitial in bulk MgO}
	To conclude this contribution we investigate the diffusion of a stochiometric hexainterstitial in MgO, whose lattice
	has space group	$Fm\bar{3}m$ with point group $\mathcal{O}_h$. In a cubic supercell with axes aligned with $\langle100\rangle$ directions we thus retain the full symmetry of the host lattice before the
	introduction of any defective structures. Connectivity graphs are constructed with vertices colored to indicate specie. Due to the high degree of symmetry,
	the maximally reduced state space contained only four states, whilst the state space irreducible under translation contained 56 states.
	\texttt{TAMMBER} was initialized in a relatively high energy state (red circle) that was found by Uberuaga
	\textit{et al.}\cite{uberuaga2005} to possess very low migration barriers; a much lower energy state was rapidly found, upon which sampling was subsequently concentrated, demonstrating the ability of \texttt{TAMMBER} to act as a massively parallel global minimum search routine.
	Using the same modified Buckingham potential as in that work, \texttt{TAMMBER} was run for 8 hours on 144 cores
	targeting a temperature range of 500-1000K. Whilst we find the very low migration barriers (<0.1eV) in agreement with
	\citep{uberuaga2005}, the full diffusion tensor was found to be isotropic ${\rm D}_1={\rm D}_2={\rm D}_3$, with a
	characteristic activation energy that converged to 0.887eV at the lower temperatures, which slowly increases
	at higher temperatures, as found for the above trimer example.
	Analyzing the dominant pathways with temperature reveals that this activation barrier corresponds to self-migration of
	the lowest energy state, with little effect of correlation on the diffusion paths; analyzing the state-by-state contribution to the uncorrelated diffusivity, shown in Fig.\ \ref{MgO}, we see that at higher temperatures the self-migration of other states have an increasingly large	contribution. $\langle\tau_{res}\rangle$ as a function of temperature are also shown in Fig.\ \ref{MgO}. The Monte Carlo procedure indicated a high degree of convergence, with $\delta\mathcal{R}<0.001$ over all temperatures considered.
\smhead{Discussion}
	We have presented a fully automatable and efficient method to evaluate the transport tensors resulting from the arbitrarily complex diffusion processes of
	crystal defects, with a well defined convergence criteria based on quantitative measures of sampling uncertainty combined with a Monte Carlo procedure to
	sample the admissable diffusion tensors consistent with sampling uncertainty and detailed balance. The method was demonstrated on a surface trimer in in
	tungsten and a hex-interstitial in magnesium oxide. By effectively eliminating user input beyond the seeding of some initial state(s), the presented approach
	demonstrates sufficient computational and critically \textit{end user} efficiencies to extend the phenomenological reach of high throughput computations to
	point defect kinetics. Future work will exploit these efficiencies to analyze defect transport over a wide range of material systems, and the influence of
	breaking detailed balance through external driving forces, giving nonzero limits for the drift vector $\bm\mu$. %\comment{We note that the diffusion tensors are defined in the limit of infinitely long trajectories; intermediate time dynamics will require the construction of more complex mesoscale models	such as reaction-diffusion equations. FINAL WORDS}
\smhead{Acknowledgements}
TDS gratefully recognizes support from the Agence Nationale de Recherche, via the MEMOPAS project ANR-19-CE46-0006-1.
This work was granted access to the HPC resources of IDRIS under the allocations AP010910718 and A0070910965 attributed by GENCI.
Work at Los Alamos National Laboratory was supported by the U. S. Department of Energy, Office of Nuclear Energy and Office of Science, Office of Advanced Scientific Computing Research through the Scientific Discovery through Advanced Computing (SciDAC) project on Fission Gas Behavior. Los Alamos National Laboratory is operated by Triad National Security LLC, for the National Nuclear Security administration of the U.S. DOE under Contract No. 89233218CNA0000001.
\smhead{Competing Interests}
The Authors declare no Competing Financial or Non-Financial Interests.
\smhead{Contributions}
TDS and DP designed the research program. TDS derived the theoretical results, designed the Monte Carlo procedure, and ran the simulations. TDS produced an initial manuscript, which was then discussed and refined with DP.
\smhead{Data Availability}
The datasets generated during and analysed during the current study are available from the corresponding author on reasonable request.
\smhead{Methods}
\small{
	\textbf{Sampling procedure}\\
	Through the use of \texttt{NEB} calculations\cite{NEB} and transition state theory\cite{Kramers} \texttt{TAMMBER} constructs a transition matrix ${\bf K}$ of rank equal to the number of discovered states, giving a continuous time Markov chain $\dot{\bf P}(t) = \left[{\bf K}-{\bf K}^{tot}\right]{\bf P}(t)$.
	Exploiting the known Poissionian distribution of exit times from a suitably thermalized basin\cite{lebris2012},
	a Bayesian likelihood (and thus posterior distribution) for the $\{\mathrm{ k}_i^u\}$ was derived using parallel trajectory data obtained through a modified temperature
	accelerated dynamics method\cite{TAD}. Continuous self-optimization was acheived by calculating the derivative of each $\mathrm{ k}_i^u$ with respect to
	additional computational work, allowing the degree of temperature acceleration to be statewise optimized and the massively parallel sampling effort distributed across states
	to differentially maximize a key measure of model validity, the expected residence time before absorbtion\cite{swinburne2018b}, equation (\ref{tau}).
	Typically, a simulation starts with one state then the rank of ${\bf P},{\bf K}$ and ${\bf K}^{tot}$ increases as states are discovered. If ${\bf P}_0$ is
	fixed to be unity for the initial state and zero otherwise, i.e. $[{\bf P}_0]_j=\delta_{ij}$ for an initial state $i$, $\langle\tau_\mathrm{ res}\rangle$ is
	monotonically increasing with sampling effort, a key consequence of the estimation procedure for $\{\mathrm{ k}^u_i\}$.
	An implementation of this method is available as an open source code\cite{tammber}, whose near-ideal parallel efficiency has been demonstrated on massively
	parallel resources employing 1000 to more than 80,000 cores.\\
	\newline
	\textbf{Isomorphically compressed representation}\\
	Isomorphic configurations are identified in two ways- the connectivity graph for every state is duplicated and reindexed into McKay's pseudounique canonical
	order\cite{NAUTY}, which is identical for isomorphic states. Alternatively, the \texttt{VF2} graph matching routine\cite{cordella2001} is applied to a single state to
	find self-isomorphisms. To determine a given reindexing $\mathcal{R}_{ij}$ between two isomorphic states $i,j$ we first find the mappings $\mathcal{R}_{ic}$
	and $\mathcal{R}_{jc}$ to the McKay canonical order then obtain $\mathcal{R}_{ij} = \mathcal{R}_{ic}\mathcal{R}^{-1}_{cj}$.

	As isomorphisms will be with respect to the simulation supercell, not the host crystal structure, the relevant point group is
	$\bar{\mathcal{G}}=\mathcal{W}\cap\mathcal{G}\in\mathcal{O}_h$, the intersection of the point group of $\mathcal{G}$ and the supercell point group
	$\mathcal{W}$, where $\mathcal{W}$ is a subgroup of the cubic group $\mathcal{O}_h$. Unlike an arbitrary element of $\mathcal{G}$ or $\mathcal{W}$, any
	element of $\bar{\mathcal{G}}$ is guaranteed to leave both the perfect lattice and supercell unchanged up to a translation and reindexing. We then iterate
	through all 48 members of $\mathcal{O}_h$, applying the point transform and applying a constant displacement such that the first indexed atom from each
	configuration	are minimum image coincident. We then check for minimium image coincidence atom-by-atom, rejecting each candidate member of $\mathcal{O}_h$ at
	the first failure.
}
\bibliography{bibliography.bib}

\begin{thebibliography}{51}%
\makeatletter
\providecommand \@ifxundefined [1]{%
 \@ifx{#1\undefined}
}%
\providecommand \@ifnum [1]{%
 \ifnum #1\expandafter \@firstoftwo
 \else \expandafter \@secondoftwo
 \fi
}%
\providecommand \@ifx [1]{%
 \ifx #1\expandafter \@firstoftwo
 \else \expandafter \@secondoftwo
 \fi
}%
\providecommand \natexlab [1]{#1}%
\providecommand \enquote  [1]{``#1''}%
\providecommand \bibnamefont  [1]{#1}%
\providecommand \bibfnamefont [1]{#1}%
\providecommand \citenamefont [1]{#1}%
\providecommand \href@noop [0]{\@secondoftwo}%
\providecommand \href [0]{\begingroup \@sanitize@url \@href}%
\providecommand \@href[1]{\@@startlink{#1}\@@href}%
\providecommand \@@href[1]{\endgroup#1\@@endlink}%
\providecommand \@sanitize@url [0]{\catcode `\\12\catcode `\$12\catcode
  `\&12\catcode `\#12\catcode `\^12\catcode `\_12\catcode `\%12\relax}%
\providecommand \@@startlink[1]{}%
\providecommand \@@endlink[0]{}%
\providecommand \url  [0]{\begingroup\@sanitize@url \@url }%
\providecommand \@url [1]{\endgroup\@href {#1}{\urlprefix }}%
\providecommand \urlprefix  [0]{URL }%
\providecommand \Eprint [0]{\href }%
\providecommand \doibase [0]{http://dx.doi.org/}%
\providecommand \selectlanguage [0]{\@gobble}%
\providecommand \bibinfo  [0]{\@secondoftwo}%
\providecommand \bibfield  [0]{\@secondoftwo}%
\providecommand \translation [1]{[#1]}%
\providecommand \BibitemOpen [0]{}%
\providecommand \bibitemStop [0]{}%
\providecommand \bibitemNoStop [0]{.\EOS\space}%
\providecommand \EOS [0]{\spacefactor3000\relax}%
\providecommand \BibitemShut  [1]{\csname bibitem#1\endcsname}%
\let\auto@bib@innerbib\@empty
%</preamble>
\bibitem [{\citenamefont {Fu}\ \emph {et~al.}(2005)\citenamefont {Fu},
  \citenamefont {Dalla~Torre}, \citenamefont {Willaime}, \citenamefont
  {Bocquet},\ and\ \citenamefont {Barbu}}]{fu2005}%
  \BibitemOpen
  \bibfield  {author} {\bibinfo {author} {\bibfnamefont {C.-C.}\ \bibnamefont
  {Fu}}, \bibinfo {author} {\bibfnamefont {J.}~\bibnamefont {Dalla~Torre}},
  \bibinfo {author} {\bibfnamefont {F.}~\bibnamefont {Willaime}}, \bibinfo
  {author} {\bibfnamefont {J.-L.}\ \bibnamefont {Bocquet}}, \ and\ \bibinfo
  {author} {\bibfnamefont {A.}~\bibnamefont {Barbu}},\ }\href@noop {}
  {\bibfield  {journal} {\bibinfo  {journal} {Nature materials}\ }\textbf
  {\bibinfo {volume} {4}},\ \bibinfo {pages} {68} (\bibinfo {year}
  {2005})}\BibitemShut {NoStop}%
\bibitem [{\citenamefont {Sefta}\ \emph {et~al.}(2013)\citenamefont {Sefta},
  \citenamefont {Hammond}, \citenamefont {Juslin},\ and\ \citenamefont
  {Wirth}}]{sefta2013}%
  \BibitemOpen
  \bibfield  {author} {\bibinfo {author} {\bibfnamefont {F.}~\bibnamefont
  {Sefta}}, \bibinfo {author} {\bibfnamefont {K.~D.}\ \bibnamefont {Hammond}},
  \bibinfo {author} {\bibfnamefont {N.}~\bibnamefont {Juslin}}, \ and\ \bibinfo
  {author} {\bibfnamefont {B.~D.}\ \bibnamefont {Wirth}},\ }\href@noop {}
  {\bibfield  {journal} {\bibinfo  {journal} {Nuclear Fusion}\ }\textbf
  {\bibinfo {volume} {53}},\ \bibinfo {pages} {073015} (\bibinfo {year}
  {2013})}\BibitemShut {NoStop}%
\bibitem [{\citenamefont {N{\o}rskov}\ \emph {et~al.}(2009)\citenamefont
  {N{\o}rskov}, \citenamefont {Bligaard}, \citenamefont {Rossmeisl},\ and\
  \citenamefont {Christensen}}]{norskov2009}%
  \BibitemOpen
  \bibfield  {author} {\bibinfo {author} {\bibfnamefont {J.~K.}\ \bibnamefont
  {N{\o}rskov}}, \bibinfo {author} {\bibfnamefont {T.}~\bibnamefont
  {Bligaard}}, \bibinfo {author} {\bibfnamefont {J.}~\bibnamefont {Rossmeisl}},
  \ and\ \bibinfo {author} {\bibfnamefont {C.~H.}\ \bibnamefont
  {Christensen}},\ }\href@noop {} {\bibfield  {journal} {\bibinfo  {journal}
  {Nature chemistry}\ }\textbf {\bibinfo {volume} {1}},\ \bibinfo {pages} {37}
  (\bibinfo {year} {2009})}\BibitemShut {NoStop}%
\bibitem [{\citenamefont {B{\"o}ller}\ \emph {et~al.}(2019)\citenamefont
  {B{\"o}ller}, \citenamefont {Durner},\ and\ \citenamefont
  {Wintterlin}}]{boller2019}%
  \BibitemOpen
  \bibfield  {author} {\bibinfo {author} {\bibfnamefont {B.}~\bibnamefont
  {B{\"o}ller}}, \bibinfo {author} {\bibfnamefont {K.~M.}\ \bibnamefont
  {Durner}}, \ and\ \bibinfo {author} {\bibfnamefont {J.}~\bibnamefont
  {Wintterlin}},\ }\href@noop {} {\bibfield  {journal} {\bibinfo  {journal}
  {Nature Catalysis}\ }\textbf {\bibinfo {volume} {2}},\ \bibinfo {pages}
  {1027} (\bibinfo {year} {2019})}\BibitemShut {NoStop}%
\bibitem [{\citenamefont {Osetsky}\ \emph {et~al.}(2018)\citenamefont
  {Osetsky}, \citenamefont {Beland}, \citenamefont {Barashev},\ and\
  \citenamefont {Zhang}}]{osetsky2018}%
  \BibitemOpen
  \bibfield  {author} {\bibinfo {author} {\bibfnamefont {Y.~N.}\ \bibnamefont
  {Osetsky}}, \bibinfo {author} {\bibfnamefont {L.~K.}\ \bibnamefont {Beland}},
  \bibinfo {author} {\bibfnamefont {A.~V.}\ \bibnamefont {Barashev}}, \ and\
  \bibinfo {author} {\bibfnamefont {Y.}~\bibnamefont {Zhang}},\ }\href@noop {}
  {\bibfield  {journal} {\bibinfo  {journal} {Current Opinion in Solid State
  and Materials Science}\ }\textbf {\bibinfo {volume} {22}},\ \bibinfo {pages}
  {65} (\bibinfo {year} {2018})}\BibitemShut {NoStop}%
\bibitem [{\citenamefont {Sorensen}\ \emph {et~al.}(2000)\citenamefont
  {Sorensen}, \citenamefont {Mishin},\ and\ \citenamefont
  {Voter}}]{sorensen2000}%
  \BibitemOpen
  \bibfield  {author} {\bibinfo {author} {\bibfnamefont {M.~R.}\ \bibnamefont
  {Sorensen}}, \bibinfo {author} {\bibfnamefont {Y.}~\bibnamefont {Mishin}}, \
  and\ \bibinfo {author} {\bibfnamefont {A.~F.}\ \bibnamefont {Voter}},\ }\href
  {\doibase 10.1103/PhysRevB.62.3658} {\bibfield  {journal} {\bibinfo
  {journal} {Phys. Rev. B}\ }\textbf {\bibinfo {volume} {62}},\ \bibinfo
  {pages} {3658} (\bibinfo {year} {2000})}\BibitemShut {NoStop}%
\bibitem [{\citenamefont {Uberuaga}\ \emph {et~al.}(2005)\citenamefont
  {Uberuaga}, \citenamefont {Smith}, \citenamefont {Cleave}, \citenamefont
  {Henkelman}, \citenamefont {Grimes}, \citenamefont {Voter},\ and\
  \citenamefont {Sickafus}}]{uberuaga2005}%
  \BibitemOpen
  \bibfield  {author} {\bibinfo {author} {\bibfnamefont {B.}~\bibnamefont
  {Uberuaga}}, \bibinfo {author} {\bibfnamefont {R.}~\bibnamefont {Smith}},
  \bibinfo {author} {\bibfnamefont {A.}~\bibnamefont {Cleave}}, \bibinfo
  {author} {\bibfnamefont {G.}~\bibnamefont {Henkelman}}, \bibinfo {author}
  {\bibfnamefont {R.}~\bibnamefont {Grimes}}, \bibinfo {author} {\bibfnamefont
  {A.}~\bibnamefont {Voter}}, \ and\ \bibinfo {author} {\bibfnamefont
  {K.}~\bibnamefont {Sickafus}},\ }\href@noop {} {\bibfield  {journal}
  {\bibinfo  {journal} {Nuclear Instruments and Methods in Physics Research
  Section B: Beam Interactions with Materials and Atoms}\ }\textbf {\bibinfo
  {volume} {228}},\ \bibinfo {pages} {260} (\bibinfo {year}
  {2005})}\BibitemShut {NoStop}%
\bibitem [{\citenamefont {Uberuaga}\ \emph {et~al.}(2007)\citenamefont
  {Uberuaga}, \citenamefont {Hoagland}, \citenamefont {Voter},\ and\
  \citenamefont {Valone}}]{uberuaga2007}%
  \BibitemOpen
  \bibfield  {author} {\bibinfo {author} {\bibfnamefont {B.}~\bibnamefont
  {Uberuaga}}, \bibinfo {author} {\bibfnamefont {R.}~\bibnamefont {Hoagland}},
  \bibinfo {author} {\bibfnamefont {A.}~\bibnamefont {Voter}}, \ and\ \bibinfo
  {author} {\bibfnamefont {S.}~\bibnamefont {Valone}},\ }\href@noop {}
  {\bibfield  {journal} {\bibinfo  {journal} {Physical review letters}\
  }\textbf {\bibinfo {volume} {99}},\ \bibinfo {pages} {135501} (\bibinfo
  {year} {2007})}\BibitemShut {NoStop}%
\bibitem [{\citenamefont {Perez}\ \emph {et~al.}(2009)\citenamefont {Perez},
  \citenamefont {Uberuaga}, \citenamefont {Shim}, \citenamefont {Amar},\ and\
  \citenamefont {Voter}}]{perez2009}%
  \BibitemOpen
  \bibfield  {author} {\bibinfo {author} {\bibfnamefont {D.}~\bibnamefont
  {Perez}}, \bibinfo {author} {\bibfnamefont {B.~P.}\ \bibnamefont {Uberuaga}},
  \bibinfo {author} {\bibfnamefont {Y.}~\bibnamefont {Shim}}, \bibinfo {author}
  {\bibfnamefont {J.~G.}\ \bibnamefont {Amar}}, \ and\ \bibinfo {author}
  {\bibfnamefont {A.~F.}\ \bibnamefont {Voter}},\ }\href@noop {} {\bibfield
  {journal} {\bibinfo  {journal} {Annual Reports in computational chemistry}\
  }\textbf {\bibinfo {volume} {5}},\ \bibinfo {pages} {79} (\bibinfo {year}
  {2009})}\BibitemShut {NoStop}%
\bibitem [{\citenamefont {B{\'e}land}\ \emph {et~al.}(2011)\citenamefont
  {B{\'e}land}, \citenamefont {Brommer}, \citenamefont {El-Mellouhi},
  \citenamefont {Joly},\ and\ \citenamefont {Mousseau}}]{beland2011}%
  \BibitemOpen
  \bibfield  {author} {\bibinfo {author} {\bibfnamefont {L.~K.}\ \bibnamefont
  {B{\'e}land}}, \bibinfo {author} {\bibfnamefont {P.}~\bibnamefont {Brommer}},
  \bibinfo {author} {\bibfnamefont {F.}~\bibnamefont {El-Mellouhi}}, \bibinfo
  {author} {\bibfnamefont {J.-F.}\ \bibnamefont {Joly}}, \ and\ \bibinfo
  {author} {\bibfnamefont {N.}~\bibnamefont {Mousseau}},\ }\href@noop {}
  {\bibfield  {journal} {\bibinfo  {journal} {Physical Review E}\ }\textbf
  {\bibinfo {volume} {84}},\ \bibinfo {pages} {046704} (\bibinfo {year}
  {2011})}\BibitemShut {NoStop}%
\bibitem [{\citenamefont {Dezerald}\ \emph {et~al.}(2015)\citenamefont
  {Dezerald}, \citenamefont {Proville}, \citenamefont {Ventelon}, \citenamefont
  {Willaime},\ and\ \citenamefont {Rodney}}]{dezerald2015first}%
  \BibitemOpen
  \bibfield  {author} {\bibinfo {author} {\bibfnamefont {L.}~\bibnamefont
  {Dezerald}}, \bibinfo {author} {\bibfnamefont {L.}~\bibnamefont {Proville}},
  \bibinfo {author} {\bibfnamefont {L.}~\bibnamefont {Ventelon}}, \bibinfo
  {author} {\bibfnamefont {F.}~\bibnamefont {Willaime}}, \ and\ \bibinfo
  {author} {\bibfnamefont {D.}~\bibnamefont {Rodney}},\ }\href@noop {}
  {\bibfield  {journal} {\bibinfo  {journal} {Physical Review B}\ }\textbf
  {\bibinfo {volume} {91}},\ \bibinfo {pages} {094105} (\bibinfo {year}
  {2015})}\BibitemShut {NoStop}%
\bibitem [{\citenamefont {Swinburne}\ \emph {et~al.}(2016)\citenamefont
  {Swinburne}, \citenamefont {Arakawa}, \citenamefont {Mori}, \citenamefont
  {Yasuda}, \citenamefont {Isshiki}, \citenamefont {Mimura}, \citenamefont
  {Uchikoshi},\ and\ \citenamefont {Dudarev}}]{swinburne2016}%
  \BibitemOpen
  \bibfield  {author} {\bibinfo {author} {\bibfnamefont {T.~D.}\ \bibnamefont
  {Swinburne}}, \bibinfo {author} {\bibfnamefont {K.}~\bibnamefont {Arakawa}},
  \bibinfo {author} {\bibfnamefont {H.}~\bibnamefont {Mori}}, \bibinfo {author}
  {\bibfnamefont {H.}~\bibnamefont {Yasuda}}, \bibinfo {author} {\bibfnamefont
  {M.}~\bibnamefont {Isshiki}}, \bibinfo {author} {\bibfnamefont
  {K.}~\bibnamefont {Mimura}}, \bibinfo {author} {\bibfnamefont
  {M.}~\bibnamefont {Uchikoshi}}, \ and\ \bibinfo {author} {\bibfnamefont
  {S.~L.}\ \bibnamefont {Dudarev}},\ }\href@noop {} {\bibfield  {journal}
  {\bibinfo  {journal} {Scientific Reports}\ }\textbf {\bibinfo {volume} {6}}
  (\bibinfo {year} {2016})}\BibitemShut {NoStop}%
\bibitem [{\citenamefont {Alexander}\ \emph {et~al.}(2016)\citenamefont
  {Alexander}, \citenamefont {Marinica}, \citenamefont {Proville},
  \citenamefont {Willaime}, \citenamefont {Arakawa}, \citenamefont {Gilbert},\
  and\ \citenamefont {Dudarev}}]{alexander2016}%
  \BibitemOpen
  \bibfield  {author} {\bibinfo {author} {\bibfnamefont {R.}~\bibnamefont
  {Alexander}}, \bibinfo {author} {\bibfnamefont {M.-C.}\ \bibnamefont
  {Marinica}}, \bibinfo {author} {\bibfnamefont {L.}~\bibnamefont {Proville}},
  \bibinfo {author} {\bibfnamefont {F.}~\bibnamefont {Willaime}}, \bibinfo
  {author} {\bibfnamefont {K.}~\bibnamefont {Arakawa}}, \bibinfo {author}
  {\bibfnamefont {M.}~\bibnamefont {Gilbert}}, \ and\ \bibinfo {author}
  {\bibfnamefont {S.}~\bibnamefont {Dudarev}},\ }\href@noop {} {\bibfield
  {journal} {\bibinfo  {journal} {Physical Review B}\ }\textbf {\bibinfo
  {volume} {94}},\ \bibinfo {pages} {024103} (\bibinfo {year}
  {2016})}\BibitemShut {NoStop}%
\bibitem [{\citenamefont {Voter}(1997)}]{voter1997}%
  \BibitemOpen
  \bibfield  {author} {\bibinfo {author} {\bibfnamefont {A.~F.}\ \bibnamefont
  {Voter}},\ }\href@noop {} {\bibfield  {journal} {\bibinfo  {journal}
  {Physical Review Letters}\ }\textbf {\bibinfo {volume} {78}},\ \bibinfo
  {pages} {3908} (\bibinfo {year} {1997})}\BibitemShut {NoStop}%
\bibitem [{\citenamefont {Voter}(1998)}]{voter1998parallel}%
  \BibitemOpen
  \bibfield  {author} {\bibinfo {author} {\bibfnamefont {A.~F.}\ \bibnamefont
  {Voter}},\ }\href@noop {} {\bibfield  {journal} {\bibinfo  {journal}
  {Physical Review B}\ }\textbf {\bibinfo {volume} {57}},\ \bibinfo {pages}
  {R13985} (\bibinfo {year} {1998})}\BibitemShut {NoStop}%
\bibitem [{\citenamefont {Sorensen}\ and\ \citenamefont {Voter}(2000)}]{TAD}%
  \BibitemOpen
  \bibfield  {author} {\bibinfo {author} {\bibfnamefont {M.}~\bibnamefont
  {Sorensen}}\ and\ \bibinfo {author} {\bibfnamefont {A.}~\bibnamefont
  {Voter}},\ }\href@noop {} {\bibfield  {journal} {\bibinfo  {journal} {The
  Journal of Chemical Physics}\ }\textbf {\bibinfo {volume} {112}},\ \bibinfo
  {pages} {9599} (\bibinfo {year} {2000})}\BibitemShut {NoStop}%
\bibitem [{\citenamefont {Chatterjee}\ and\ \citenamefont
  {Bhattacharya}(2015)}]{chatterjee2015uncertainty}%
  \BibitemOpen
  \bibfield  {author} {\bibinfo {author} {\bibfnamefont {A.}~\bibnamefont
  {Chatterjee}}\ and\ \bibinfo {author} {\bibfnamefont {S.}~\bibnamefont
  {Bhattacharya}},\ }\href@noop {} {\bibfield  {journal} {\bibinfo  {journal}
  {The Journal of Chemical Physics}\ }\textbf {\bibinfo {volume} {143}},\
  \bibinfo {pages} {114109} (\bibinfo {year} {2015})}\BibitemShut {NoStop}%
\bibitem [{\citenamefont {Chill}\ and\ \citenamefont
  {Henkelman}(2014)}]{chill2014}%
  \BibitemOpen
  \bibfield  {author} {\bibinfo {author} {\bibfnamefont {S.~T.}\ \bibnamefont
  {Chill}}\ and\ \bibinfo {author} {\bibfnamefont {G.}~\bibnamefont
  {Henkelman}},\ }\href@noop {} {\bibfield  {journal} {\bibinfo  {journal} {The
  Journal of chemical physics}\ }\textbf {\bibinfo {volume} {140}},\ \bibinfo
  {pages} {214110} (\bibinfo {year} {2014})}\BibitemShut {NoStop}%
\bibitem [{\citenamefont {Henkelman}\ and\ \citenamefont
  {J{\'o}nsson}(1999)}]{dimer}%
  \BibitemOpen
  \bibfield  {author} {\bibinfo {author} {\bibfnamefont {G.}~\bibnamefont
  {Henkelman}}\ and\ \bibinfo {author} {\bibfnamefont {H.}~\bibnamefont
  {J{\'o}nsson}},\ }\href@noop {} {\bibfield  {journal} {\bibinfo  {journal}
  {The Journal of chemical physics}\ }\textbf {\bibinfo {volume} {111}},\
  \bibinfo {pages} {7010} (\bibinfo {year} {1999})}\BibitemShut {NoStop}%
\bibitem [{\citenamefont {Wales}(2002)}]{wales2002discrete}%
  \BibitemOpen
  \bibfield  {author} {\bibinfo {author} {\bibfnamefont {D.~J.}\ \bibnamefont
  {Wales}},\ }\href@noop {} {\bibfield  {journal} {\bibinfo  {journal}
  {Molecular physics}\ }\textbf {\bibinfo {volume} {100}},\ \bibinfo {pages}
  {3285} (\bibinfo {year} {2002})}\BibitemShut {NoStop}%
\bibitem [{\citenamefont {Wales}(2003)}]{wales_energy_2003}%
  \BibitemOpen
  \bibfield  {author} {\bibinfo {author} {\bibfnamefont {D.~J.}\ \bibnamefont
  {Wales}},\ }\href@noop {} {\emph {\bibinfo {title} {Energy Landscapes}}},\
  edited by\ \bibinfo {editor} {\bibfnamefont {C.~U.}\ \bibnamefont {Press}}\
  (\bibinfo  {publisher} {Cambridge},\ \bibinfo {year} {2003})\BibitemShut
  {NoStop}%
\bibitem [{\citenamefont {Le~Bris}\ \emph {et~al.}(2012)\citenamefont
  {Le~Bris}, \citenamefont {Lelievre}, \citenamefont {Luskin},\ and\
  \citenamefont {Perez}}]{lebris2012}%
  \BibitemOpen
  \bibfield  {author} {\bibinfo {author} {\bibfnamefont {C.}~\bibnamefont
  {Le~Bris}}, \bibinfo {author} {\bibfnamefont {T.}~\bibnamefont {Lelievre}},
  \bibinfo {author} {\bibfnamefont {M.}~\bibnamefont {Luskin}}, \ and\ \bibinfo
  {author} {\bibfnamefont {D.}~\bibnamefont {Perez}},\ }\href@noop {}
  {\bibfield  {journal} {\bibinfo  {journal} {Monte Carlo Methods and
  Applications}\ }\textbf {\bibinfo {volume} {18}},\ \bibinfo {pages} {119}
  (\bibinfo {year} {2012})}\BibitemShut {NoStop}%
\bibitem [{\citenamefont {Leli{\`e}vre}(2018)}]{Lelièvre2018}%
  \BibitemOpen
  \bibfield  {author} {\bibinfo {author} {\bibfnamefont {T.}~\bibnamefont
  {Leli{\`e}vre}},\ }\enquote {\bibinfo {title} {Mathematical foundations of
  accelerated molecular dynamics methods},}\ in\ \href {\doibase
  10.1007/978-3-319-42913-7_27-1} {\emph {\bibinfo {booktitle} {Handbook of
  Materials Modeling : Methods: Theory and Modeling}}},\ \bibinfo {editor}
  {edited by\ \bibinfo {editor} {\bibfnamefont {W.}~\bibnamefont {Andreoni}}\
  and\ \bibinfo {editor} {\bibfnamefont {S.}~\bibnamefont {Yip}}}\ (\bibinfo
  {publisher} {Springer International Publishing},\ \bibinfo {address} {Cham},\
  \bibinfo {year} {2018})\ pp.\ \bibinfo {pages} {1--32}\BibitemShut {NoStop}%
\bibitem [{\citenamefont {Henkelman}(2017)}]{henkelman2017}%
  \BibitemOpen
  \bibfield  {author} {\bibinfo {author} {\bibfnamefont {G.}~\bibnamefont
  {Henkelman}},\ }\href@noop {} {\bibfield  {journal} {\bibinfo  {journal}
  {Annual Review of Materials Research}\ } (\bibinfo {year}
  {2017})}\BibitemShut {NoStop}%
\bibitem [{\citenamefont {Aristoff}\ \emph {et~al.}(2016)\citenamefont
  {Aristoff}, \citenamefont {Chill},\ and\ \citenamefont
  {Simpson}}]{aristoff2016}%
  \BibitemOpen
  \bibfield  {author} {\bibinfo {author} {\bibfnamefont {D.}~\bibnamefont
  {Aristoff}}, \bibinfo {author} {\bibfnamefont {S.}~\bibnamefont {Chill}}, \
  and\ \bibinfo {author} {\bibfnamefont {G.}~\bibnamefont {Simpson}},\
  }\href@noop {} {\bibfield  {journal} {\bibinfo  {journal} {Communications in
  Applied Mathematics and Computational Science}\ }\textbf {\bibinfo {volume}
  {11}},\ \bibinfo {pages} {171} (\bibinfo {year} {2016})}\BibitemShut
  {NoStop}%
\bibitem [{\citenamefont {Swinburne}\ and\ \citenamefont
  {Perez}(2018{\natexlab{a}})}]{swinburne2018b}%
  \BibitemOpen
  \bibfield  {author} {\bibinfo {author} {\bibfnamefont {T.~D.}\ \bibnamefont
  {Swinburne}}\ and\ \bibinfo {author} {\bibfnamefont {D.}~\bibnamefont
  {Perez}},\ }\href {\doibase 10.1103/PhysRevMaterials.2.053802} {\bibfield
  {journal} {\bibinfo  {journal} {Phys. Rev. Materials}\ }\textbf {\bibinfo
  {volume} {2}},\ \bibinfo {pages} {053802} (\bibinfo {year}
  {2018}{\natexlab{a}})}\BibitemShut {NoStop}%
\bibitem [{\citenamefont {Bart{\'o}k}\ \emph {et~al.}(2017)\citenamefont
  {Bart{\'o}k}, \citenamefont {De}, \citenamefont {Poelking}, \citenamefont
  {Bernstein}, \citenamefont {Kermode}, \citenamefont {Cs{\'a}nyi},\ and\
  \citenamefont {Ceriotti}}]{bartok2017}%
  \BibitemOpen
  \bibfield  {author} {\bibinfo {author} {\bibfnamefont {A.~P.}\ \bibnamefont
  {Bart{\'o}k}}, \bibinfo {author} {\bibfnamefont {S.}~\bibnamefont {De}},
  \bibinfo {author} {\bibfnamefont {C.}~\bibnamefont {Poelking}}, \bibinfo
  {author} {\bibfnamefont {N.}~\bibnamefont {Bernstein}}, \bibinfo {author}
  {\bibfnamefont {J.~R.}\ \bibnamefont {Kermode}}, \bibinfo {author}
  {\bibfnamefont {G.}~\bibnamefont {Cs{\'a}nyi}}, \ and\ \bibinfo {author}
  {\bibfnamefont {M.}~\bibnamefont {Ceriotti}},\ }\href@noop {} {\bibfield
  {journal} {\bibinfo  {journal} {Science advances}\ }\textbf {\bibinfo
  {volume} {3}},\ \bibinfo {pages} {e1701816} (\bibinfo {year}
  {2017})}\BibitemShut {NoStop}%
\bibitem [{\citenamefont {Marinica}(2019)}]{cosmin2019}%
  \BibitemOpen
  \bibfield  {author} {\bibinfo {author} {\bibfnamefont {M.}~\bibnamefont
  {Marinica}},\ }\href {\doibase
  https://doi.org/10.1016/j.commatsci.2019.04.043} {\bibfield  {journal}
  {\bibinfo  {journal} {Computational Materials Science}\ }\textbf {\bibinfo
  {volume} {166}},\ \bibinfo {pages} {200 } (\bibinfo {year}
  {2019})}\BibitemShut {NoStop}%
\bibitem [{\citenamefont {Shapeev}(2016)}]{shapeev2016moment}%
  \BibitemOpen
  \bibfield  {author} {\bibinfo {author} {\bibfnamefont {A.~V.}\ \bibnamefont
  {Shapeev}},\ }\href@noop {} {\bibfield  {journal} {\bibinfo  {journal}
  {Multiscale Modeling \& Simulation}\ }\textbf {\bibinfo {volume} {14}},\
  \bibinfo {pages} {1153} (\bibinfo {year} {2016})}\BibitemShut {NoStop}%
\bibitem [{\citenamefont {Boulougouris}\ and\ \citenamefont
  {Frenkel}(2005)}]{boulougouris2005monte}%
  \BibitemOpen
  \bibfield  {author} {\bibinfo {author} {\bibfnamefont {G.~C.}\ \bibnamefont
  {Boulougouris}}\ and\ \bibinfo {author} {\bibfnamefont {D.}~\bibnamefont
  {Frenkel}},\ }\href@noop {} {\bibfield  {journal} {\bibinfo  {journal}
  {Journal of chemical theory and computation}\ }\textbf {\bibinfo {volume}
  {1}},\ \bibinfo {pages} {389} (\bibinfo {year} {2005})}\BibitemShut {NoStop}%
\bibitem [{\citenamefont {Bhoutekar}\ \emph {et~al.}(2017)\citenamefont
  {Bhoutekar}, \citenamefont {Ghosh}, \citenamefont {Bhattacharya},\ and\
  \citenamefont {Chatterjee}}]{bhoutekar2017new}%
  \BibitemOpen
  \bibfield  {author} {\bibinfo {author} {\bibfnamefont {A.}~\bibnamefont
  {Bhoutekar}}, \bibinfo {author} {\bibfnamefont {S.}~\bibnamefont {Ghosh}},
  \bibinfo {author} {\bibfnamefont {S.}~\bibnamefont {Bhattacharya}}, \ and\
  \bibinfo {author} {\bibfnamefont {A.}~\bibnamefont {Chatterjee}},\
  }\href@noop {} {\bibfield  {journal} {\bibinfo  {journal} {The Journal of
  Chemical Physics}\ }\textbf {\bibinfo {volume} {147}},\ \bibinfo {pages}
  {152702} (\bibinfo {year} {2017})}\BibitemShut {NoStop}%
\bibitem [{\citenamefont {Perez}\ \emph {et~al.}(2015)\citenamefont {Perez},
  \citenamefont {Cubuk}, \citenamefont {Waterland}, \citenamefont {Kaxiras},\
  and\ \citenamefont {Voter}}]{perez2015}%
  \BibitemOpen
  \bibfield  {author} {\bibinfo {author} {\bibfnamefont {D.}~\bibnamefont
  {Perez}}, \bibinfo {author} {\bibfnamefont {E.~D.}\ \bibnamefont {Cubuk}},
  \bibinfo {author} {\bibfnamefont {A.}~\bibnamefont {Waterland}}, \bibinfo
  {author} {\bibfnamefont {E.}~\bibnamefont {Kaxiras}}, \ and\ \bibinfo
  {author} {\bibfnamefont {A.~F.}\ \bibnamefont {Voter}},\ }\href@noop {}
  {\bibfield  {journal} {\bibinfo  {journal} {Journal of chemical theory and
  computation}\ }\textbf {\bibinfo {volume} {12}},\ \bibinfo {pages} {18}
  (\bibinfo {year} {2015})}\BibitemShut {NoStop}%
\bibitem [{\citenamefont {McKay}\ and\ \citenamefont {Piperno}(2014)}]{NAUTY}%
  \BibitemOpen
  \bibfield  {author} {\bibinfo {author} {\bibfnamefont {B.~D.}\ \bibnamefont
  {McKay}}\ and\ \bibinfo {author} {\bibfnamefont {A.}~\bibnamefont
  {Piperno}},\ }\href {\doibase https://doi.org/10.1016/j.jsc.2013.09.003}
  {\bibfield  {journal} {\bibinfo  {journal} {Journal of Symbolic Computation}\
  }\textbf {\bibinfo {volume} {60}},\ \bibinfo {pages} {94 } (\bibinfo {year}
  {2014})}\BibitemShut {NoStop}%
\bibitem [{\citenamefont {Scott}(2012)}]{scott2012group}%
  \BibitemOpen
  \bibfield  {author} {\bibinfo {author} {\bibfnamefont {W.~R.}\ \bibnamefont
  {Scott}},\ }\href@noop {} {\emph {\bibinfo {title} {Group theory}}}\
  (\bibinfo  {publisher} {Courier Corporation},\ \bibinfo {year}
  {2012})\BibitemShut {NoStop}%
\bibitem [{Note1()}]{Note1}%
  \BibitemOpen
  \bibinfo {note} {Right coset are formed through
  premultiplication}\BibitemShut {NoStop}%
\bibitem [{\citenamefont {Stukowski}(2010)}]{OVITO}%
  \BibitemOpen
  \bibfield  {author} {\bibinfo {author} {\bibfnamefont {A.}~\bibnamefont
  {Stukowski}},\ }\href {http://stacks.iop.org/0965-0393/18/i=1/a=015012}
  {\bibfield  {journal} {\bibinfo  {journal} {Modelling and Simulation in
  Materials Science and Engineering}\ }\textbf {\bibinfo {volume} {18}},\
  \bibinfo {pages} {015012} (\bibinfo {year} {2010})}\BibitemShut {NoStop}%
\bibitem [{\citenamefont {Bortz}\ \emph {et~al.}(1975)\citenamefont {Bortz},
  \citenamefont {Kalos},\ and\ \citenamefont {Lebowitz}}]{bortz1975}%
  \BibitemOpen
  \bibfield  {author} {\bibinfo {author} {\bibfnamefont {A.~B.}\ \bibnamefont
  {Bortz}}, \bibinfo {author} {\bibfnamefont {M.~H.}\ \bibnamefont {Kalos}}, \
  and\ \bibinfo {author} {\bibfnamefont {J.~L.}\ \bibnamefont {Lebowitz}},\
  }\href@noop {} {\bibfield  {journal} {\bibinfo  {journal} {Journal of
  Computational Physics}\ }\textbf {\bibinfo {volume} {17}},\ \bibinfo {pages}
  {10} (\bibinfo {year} {1975})}\BibitemShut {NoStop}%
\bibitem [{\citenamefont {Trinkle}(2018)}]{Trinkle2018}%
  \BibitemOpen
  \bibfield  {author} {\bibinfo {author} {\bibfnamefont {D.~R.}\ \bibnamefont
  {Trinkle}},\ }\href@noop {} {\bibfield  {journal} {\bibinfo  {journal}
  {Physical review letters}\ }\textbf {\bibinfo {volume} {121}},\ \bibinfo
  {pages} {235901} (\bibinfo {year} {2018})}\BibitemShut {NoStop}%
\bibitem [{\citenamefont {Landman}\ and\ \citenamefont
  {Shlesinger}(1979)}]{landman1979stochastic}%
  \BibitemOpen
  \bibfield  {author} {\bibinfo {author} {\bibfnamefont {U.}~\bibnamefont
  {Landman}}\ and\ \bibinfo {author} {\bibfnamefont {M.~F.}\ \bibnamefont
  {Shlesinger}},\ }\href@noop {} {\bibfield  {journal} {\bibinfo  {journal}
  {Physical Review B}\ }\textbf {\bibinfo {volume} {19}},\ \bibinfo {pages}
  {6207} (\bibinfo {year} {1979})}\BibitemShut {NoStop}%
\bibitem [{\citenamefont {Cover}\ and\ \citenamefont
  {Thomas}(2012)}]{cover2012}%
  \BibitemOpen
  \bibfield  {author} {\bibinfo {author} {\bibfnamefont {T.~M.}\ \bibnamefont
  {Cover}}\ and\ \bibinfo {author} {\bibfnamefont {J.~A.}\ \bibnamefont
  {Thomas}},\ }\href@noop {} {\emph {\bibinfo {title} {Elements of information
  theory}}}\ (\bibinfo  {publisher} {John Wiley \& Sons},\ \bibinfo {year}
  {2012})\BibitemShut {NoStop}%
\bibitem [{Note2()}]{Note2}%
  \BibitemOpen
  \bibinfo {note} {In this case we have a separate QSD ${\protect \bm {\pi
  }}^{QSD}_s$ for each connected subnetwork $s$, which only has support on that
  subnetwork. We then calculate ${\protect \bf D}({\protect \bf k}^{u,c})$ with
  ${\protect \bf P}_0=\DOTSB \sum@ \slimits@ _s{\protect \bm {\pi }}^{QSD}_s$,
  and the limiting eigenspectrum is degenerate}\BibitemShut {NoStop}%
\bibitem [{\citenamefont {Marinica}\ \emph {et~al.}(2012)\citenamefont
  {Marinica}, \citenamefont {Willaime},\ and\ \citenamefont
  {Crocombette}}]{marinica2012}%
  \BibitemOpen
  \bibfield  {author} {\bibinfo {author} {\bibfnamefont {M.-C.}\ \bibnamefont
  {Marinica}}, \bibinfo {author} {\bibfnamefont {F.}~\bibnamefont {Willaime}},
  \ and\ \bibinfo {author} {\bibfnamefont {J.-P.}\ \bibnamefont
  {Crocombette}},\ }\href {\doibase 10.1103/PhysRevLett.108.025501} {\bibfield
  {journal} {\bibinfo  {journal} {Phys. Rev. Lett.}\ }\textbf {\bibinfo
  {volume} {108}},\ \bibinfo {pages} {025501} (\bibinfo {year}
  {2012})}\BibitemShut {NoStop}%
\bibitem [{\citenamefont {Yang}\ \emph {et~al.}(2019)\citenamefont {Yang},
  \citenamefont {Wirth}, \citenamefont {Perez},\ and\ \citenamefont
  {Voter}}]{yang2019}%
  \BibitemOpen
  \bibfield  {author} {\bibinfo {author} {\bibfnamefont {L.}~\bibnamefont
  {Yang}}, \bibinfo {author} {\bibfnamefont {B.}~\bibnamefont {Wirth}},
  \bibinfo {author} {\bibfnamefont {D.}~\bibnamefont {Perez}}, \ and\ \bibinfo
  {author} {\bibfnamefont {A.~F.}\ \bibnamefont {Voter}},\ }\href {\doibase
  https://doi.org/10.1016/j.nimb.2019.05.078} {\bibfield  {journal} {\bibinfo
  {journal} {Nuclear Instruments and Methods in Physics Research Section B:
  Beam Interactions with Materials and Atoms}\ }\textbf {\bibinfo {volume}
  {453}},\ \bibinfo {pages} {61 } (\bibinfo {year} {2019})}\BibitemShut
  {NoStop}%
\bibitem [{\citenamefont {Marinica}\ \emph {et~al.}(2013)\citenamefont
  {Marinica}, \citenamefont {Ventelon}, \citenamefont {Gilbert}, \citenamefont
  {Proville}, \citenamefont {Dudarev}, \citenamefont {Marian}, \citenamefont
  {Bencteux},\ and\ \citenamefont {Willaime}}]{marinica2013}%
  \BibitemOpen
  \bibfield  {author} {\bibinfo {author} {\bibfnamefont {M.~C.}\ \bibnamefont
  {Marinica}}, \bibinfo {author} {\bibfnamefont {L.}~\bibnamefont {Ventelon}},
  \bibinfo {author} {\bibfnamefont {M.~R.}\ \bibnamefont {Gilbert}}, \bibinfo
  {author} {\bibfnamefont {L.}~\bibnamefont {Proville}}, \bibinfo {author}
  {\bibfnamefont {S.~L.}\ \bibnamefont {Dudarev}}, \bibinfo {author}
  {\bibfnamefont {J.}~\bibnamefont {Marian}}, \bibinfo {author} {\bibfnamefont
  {G.}~\bibnamefont {Bencteux}}, \ and\ \bibinfo {author} {\bibfnamefont
  {F.}~\bibnamefont {Willaime}},\ }\href@noop {} {\bibfield  {journal}
  {\bibinfo  {journal} {Journal of Physics: Condensed Matter}\ }\textbf
  {\bibinfo {volume} {25}},\ \bibinfo {pages} {395502} (\bibinfo {year}
  {2013})}\BibitemShut {NoStop}%
\bibitem [{\citenamefont {Dijkstra}(1959)}]{dijkstra59}%
  \BibitemOpen
  \bibfield  {author} {\bibinfo {author} {\bibfnamefont {E.~W.}\ \bibnamefont
  {Dijkstra}},\ }\href@noop {} {\bibfield  {journal} {\bibinfo  {journal}
  {Numerische Math.}\ }\textbf {\bibinfo {volume} {1}},\ \bibinfo {pages} {269}
  (\bibinfo {year} {1959})}\BibitemShut {NoStop}%
\bibitem [{\citenamefont {Juslin}\ and\ \citenamefont
  {Wirth}(2013)}]{juslin2013}%
  \BibitemOpen
  \bibfield  {author} {\bibinfo {author} {\bibfnamefont {N.}~\bibnamefont
  {Juslin}}\ and\ \bibinfo {author} {\bibfnamefont {B.~D.}\ \bibnamefont
  {Wirth}},\ }\href {\doibase https://doi.org/10.1016/j.jnucmat.2012.07.023}
  {\bibfield  {journal} {\bibinfo  {journal} {Journal of Nuclear Materials}\
  }\textbf {\bibinfo {volume} {432}},\ \bibinfo {pages} {61 } (\bibinfo {year}
  {2013})}\BibitemShut {NoStop}%
\bibitem [{\citenamefont {Henkelman}\ \emph {et~al.}(2000)\citenamefont
  {Henkelman}, \citenamefont {Uberuaga},\ and\ \citenamefont {Jonsson}}]{NEB}%
  \BibitemOpen
  \bibfield  {author} {\bibinfo {author} {\bibfnamefont {G.}~\bibnamefont
  {Henkelman}}, \bibinfo {author} {\bibfnamefont {B.~P.}\ \bibnamefont
  {Uberuaga}}, \ and\ \bibinfo {author} {\bibfnamefont {H.}~\bibnamefont
  {Jonsson}},\ }\href@noop {} {\bibfield  {journal} {\bibinfo  {journal} {The
  Journal of Chemical Physics}\ }\textbf {\bibinfo {volume} {113}},\ \bibinfo
  {pages} {9901} (\bibinfo {year} {2000})}\BibitemShut {NoStop}%
\bibitem [{\citenamefont {H\"{a}nggi}\ \emph {et~al.}(1990)\citenamefont
  {H\"{a}nggi}, \citenamefont {Talkner},\ and\ \citenamefont
  {Borkovec}}]{Kramers}%
  \BibitemOpen
  \bibfield  {author} {\bibinfo {author} {\bibfnamefont {P.}~\bibnamefont
  {H\"{a}nggi}}, \bibinfo {author} {\bibfnamefont {P.}~\bibnamefont {Talkner}},
  \ and\ \bibinfo {author} {\bibfnamefont {M.}~\bibnamefont {Borkovec}},\
  }\href@noop {} {\bibfield  {journal} {\bibinfo  {journal} {Reviews of Modern
  Physics}\ }\textbf {\bibinfo {volume} {62}},\ \bibinfo {pages} {251}
  (\bibinfo {year} {1990})}\BibitemShut {NoStop}%
\bibitem [{\citenamefont {Swinburne}\ and\ \citenamefont
  {Perez}(2018{\natexlab{b}})}]{tammber}%
  \BibitemOpen
  \bibfield  {author} {\bibinfo {author} {\bibfnamefont {T.}~\bibnamefont
  {Swinburne}}\ and\ \bibinfo {author} {\bibfnamefont {D.}~\bibnamefont
  {Perez}},\ }\href {https://gitlab.com/exaalt/parsplice/tree/tammber}
  {\enquote {\bibinfo {title} {\texttt{TAMMBER} branch of \texttt{ParSplice}
  code},}\ } (\bibinfo {year} {2018}{\natexlab{b}})\BibitemShut {NoStop}%
\bibitem [{\citenamefont {Cordella}\ \emph {et~al.}(2001)\citenamefont
  {Cordella}, \citenamefont {Foggia}, \citenamefont {Sansone},\ and\
  \citenamefont {Vento}}]{cordella2001}%
  \BibitemOpen
  \bibfield  {author} {\bibinfo {author} {\bibfnamefont {L.~P.}\ \bibnamefont
  {Cordella}}, \bibinfo {author} {\bibfnamefont {P.}~\bibnamefont {Foggia}},
  \bibinfo {author} {\bibfnamefont {C.}~\bibnamefont {Sansone}}, \ and\
  \bibinfo {author} {\bibfnamefont {M.}~\bibnamefont {Vento}},\ }in\ \href@noop
  {} {\emph {\bibinfo {booktitle} {3rd IAPR-TC15 workshop on graph-based
  representations in pattern recognition}}}\ (\bibinfo {year} {2001})\ pp.\
  \bibinfo {pages} {149--159}\BibitemShut {NoStop}%
\bibitem [{Note3()}]{Note3}%
  \BibitemOpen
  \bibinfo {note} {See equation S15 of the supplementary material for the most
  direct comparison}\BibitemShut {NoStop}%
\end{thebibliography}%
\onecolumngrid
\appendix
\section{Effect of sampling in irreducible representation}
\begin{figure}[!h]
	\includegraphics[width=0.95\textwidth]{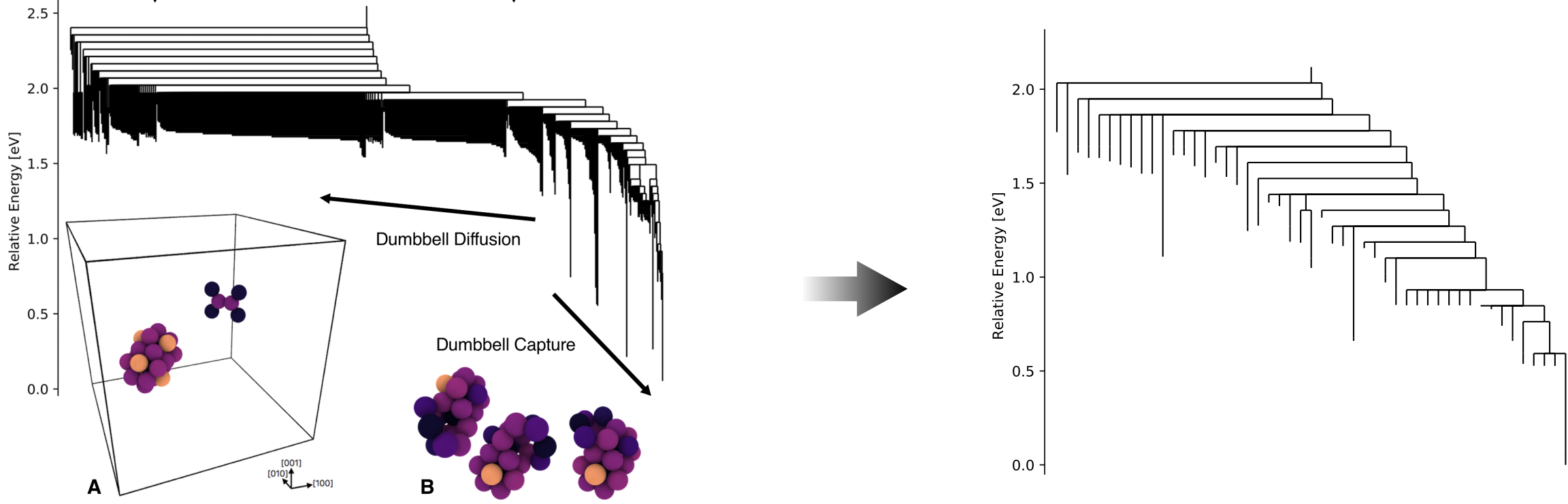}
	\includegraphics[width=0.95\textwidth]{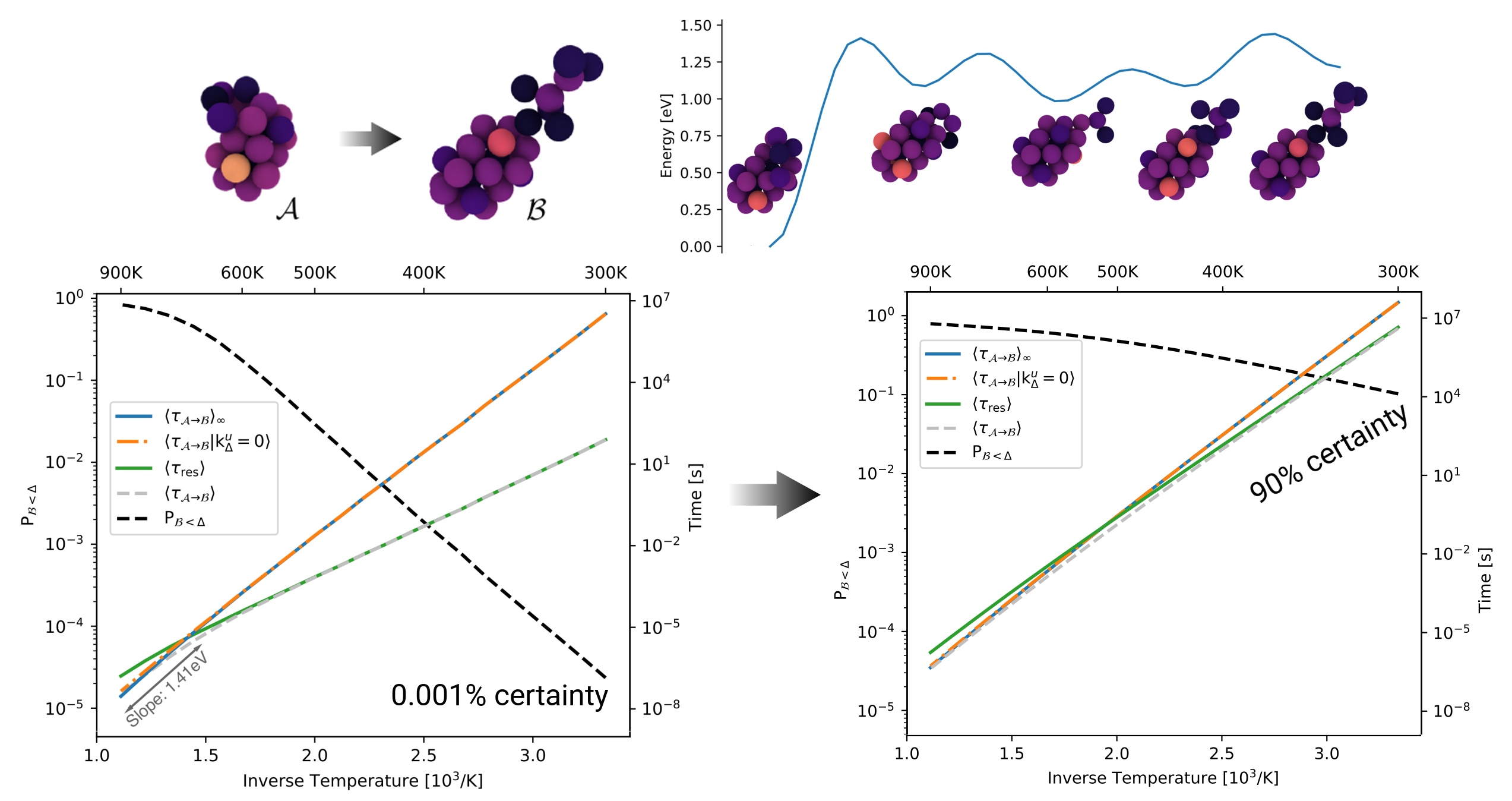}
	\caption{Above: Disconnectivity graph\cite{wales_energy_2003} for the states and transitions found by \texttt{TAMMBER} using the previously presented reducibile (left) and new irreducible (right) representation. Middle: Representative structures of the $\mathcal{A}$ and $\mathcal{B}$ basins (left) and the found minimum energy pathway (right). Below: Probability ${\rm P}_{\mathcal{B}<\Delta}$ that the system executes a $\mathcal{A\to B}$ transition before absorbtion, residence times $\langle\tau_{res}\rangle$ and various approximate projected breakup times\cite{swinburne2018b} at a range of temperatures using a Markov model built using the previously presented reducible (left) or new irreducible (right) representation. We see the irreducible representation achieves a significant compression of phase space and yields much greater certainty, as encoded by ${\rm P}_{\mathcal{B}<\Delta}$, that our model has sufficient predictive power to capture the dominant breakup mechanism.}
	\label{fig}
\end{figure}

In previous work,\cite{swinburne2018b} we used an initial version of \texttt{TAMMBER} to study the breakup of a C15-dumbell interstitial cluster in an EAM model of iron. We refer the reader to the original paper for details. We later performed an identical simulation using the new version of \texttt{TAMMBER} that compresses the state space for sampling to a representation irreducible under exchange, space group and translational symmetries, as described in the main text. The compression in the size of state space, and the resultant effect on the model quality, are summarized in figure \ref{fig}. Interestingly, despite the low degree of space group symmetry for this system, due to the presence of multiple nonlocal exchange events during sampling, the compression of exchange symmetries produced a significant efficiency improvement.

\section{Equivalence of derivatives of the generating function to pathwise averages}
We start by recalling the branching probabilty matrices for $N$ irreducible states in $D$ dimensions
\begin{align}
	[{\bf B}^{u,c}]_p&=\frac{{\rm k}^{u,c}_p}{{\rm k}^{tot,c}_p}
	,\quad
	[{\bf B}^{c}({\bm\lambda})]_{qp} \equiv
	\sum_{l\in\mathcal{C}_{qp}} \frac{{\rm k}^{c}_{l}}{{\rm k}^{tot,c}_p}
	\exp\left(-{\bm\lambda}\cdot{\bf d}_{l}\right)
	,\quad
	[{\bf B}^{c}({\bf0})]_{qp} = \frac{{\rm k}^{c}_{qp}}{{\rm k}^{tot,c}_p},\nonumber
\end{align}
In the following, we will suppress the $c$ superscript for clarity of presentation, meaning ${\bf B}^{u,c}\to{\bf B}^{u},{\bf B}^{c}\to{\bf B}$, etc.\\
The generating function with arbitrary initial condition ${\bf P}(0)$ reads
\begin{align}
	Z({\bm\lambda})&=
	{\mathbf B}^{u}\sum_n\left[{\bf B}^{}({\bm\lambda})\right]^n{\bf P}(0)=
	{\bf B}^{u}\left[\mathbb{I}-{\bf B}^{}({\bm\lambda})\right]^{-1}{\bf P}(0)=
	{\bf B}^{u}{\bf G}^{}({\bm\lambda}){\bf P}(0)
	\label{app:gf}.
\end{align}
By conservation of probabilty, it is clear that $[{\bf B}^{u}]_p + \sum_q [{\bf B}^{}({\bf0})]_{qp} = 1$,
which implies that ${\bf B}^{u}[\mathbb{I}-{\bf B}^{}({\bf0})]^{-1} = {\bf1}$ and thus $Z({\bf 0}) = {\bf1}\cdot{\bf P}(0) = 1$.\\
\newline
Consider the set $\mathcal{P}_n$ of paths $\xi=\{\xi_i\}_0^{n}$ which execute $n$ transitions between $n+1$ states before absorbtion. The probability weight $P_{\xi}$ of each path and all $n$-paths is given by
\begin{equation}
	P_{\xi} = \left[{\bf B}^{u}\right]_{\xi_n}
	\left(\prod_{l=1}^{n}
	\left[{\bf B}^{}({\bf0})\right]_{\xi_l\xi_{l-1}}\right)
	\left[{\bf P}(0)\right]_{\xi_0}
	,\quad\Rightarrow
	\sum_{\xi\in\mathcal{P}_n} P_\xi = {\bf B}^{u}\left[{\bf B}^{}({\bf0})\right]^n{\bf P}(0)
\end{equation}
The sum over of path weights over the set of all possible paths $\mathcal{P}=\cup_n\mathcal{P}_n$ can be confirmed to be unity as expected-
\begin{equation}
	\sum_{\xi\in\mathcal{P}} P_\xi = \sum_n \sum_{\xi\in\mathcal{P}_n} P_\xi = \sum_n {\bf B}^{u}\left[{\bf B}^{}({\bf0})\right]^n{\bf P}(0) =
	{\bf B}^{u}\left[\mathbb{I}-{\bf B}^{}({\bf0})\right]^{-1}{\bf P}(0) = Z({\bf 0}) = 1.
\end{equation}
The total displacement ${\bf x}_{\xi} = \sum_l{\bf d}_{\xi_l\xi_{l-1}}$ across each path can be found by simply constructing the pathwise probability as before, but replacing ${\bf B}^{}({\bf0})$ with ${\bf B}^{}({\bm\lambda})$ to yield
\begin{equation}
	P_{\xi}({\bm\lambda})
	= \left[{\bf B}^{u}\right]_{\xi_n}
	\left(\prod_{l=1}^n
	\left[{\bf B}^{}({\bf0})\right]_{\xi_l\xi_{l-1}}\right)
	%\exp\left[
	\exp^{-{\bm\lambda}\cdot\left(\sum_l{\bf d}_{\xi_l\xi_{l-1}}\right)}
	%\right]
	\left[{\bf P}(0)\right]_{\xi_0} = P_{\xi}
	\exp^{-{\bm\lambda}\cdot{\bf x}_{\xi}}
	\,\Rightarrow\,
	{\bf x}_{\xi} = -\partial_{\bm\lambda} \ln P_{\xi}({\bm\lambda}) \Big|_{\bm\lambda=\bf0}.
\end{equation}
The expected value (or first moment) of the total displacement over the set of all possible paths is therefore
\begin{equation}
	\langle{\bf x}\rangle = \sum_{\xi\in\mathcal{P}} {\bf x}_\xi P_\xi
	= -\partial_{\bm\lambda}\sum_{\xi\in\mathcal{P}} P_\xi({\bm\lambda})\Big|_{\bm\lambda=\bf0}
	= -\partial_{\bm\lambda} Z({\bm\lambda})\Big|_{\bm\lambda=\bf0}.
\end{equation}
The second moment matrix of the total displacement follows naturally-
\begin{equation}
	\langle{\bf x}\otimes{\bf x}\rangle = \sum_{\xi\in\mathcal{P}} {\bf x}_\xi\otimes{\bf x}_\xi P_\xi
	= \partial_{\bm\lambda}\otimes\partial_{\bm\lambda}\sum_{\xi\in\mathcal{P}} P_\xi({\bm\lambda})\Big|_{\bm\lambda=\bf0}
	= \partial_{\bm\lambda}\otimes\partial_{\bm\lambda} Z({\bm\lambda})\Big|_{\bm\lambda=\bf0},
\end{equation}
where $\otimes$ is the outer (or dyadic) product.
\section{Evaluating derivatives of the generating function}
Consider the following application of the chain rule, using the above definition ${\bf G}^{}({\bm\lambda})=\left[\mathbb{I}-{\bf B}^{}({\bm\lambda})\right]^{-1}$,
but writing ${\bf G}_{\bm\lambda}^{}$ instead of ${\bf G}^{}({\bm\lambda})$ for readability-
\begin{equation}
	\partial_{\lambda_\alpha}\left({\bf G}^{}\left[\mathbb{I}-{\bf B}^{}_{\bm\lambda}\right]\right)
	=
	\partial_{\lambda_\alpha}{\bf G}^{}_{\bm\lambda}\left[\mathbb{I}-{\bf B}^{}_{\bm\lambda}\right]
	-
	{\bf G}^{}_{\bm\lambda}\partial_{\lambda_\alpha}{\bf B}^{}_{\bm\lambda}
	= {\bf 0}
	\quad\Rightarrow
	\partial_{\lambda_\alpha}{\bf G}^{}_{\bm\lambda}
	=
	{\bf G}^{}_{\bm\lambda}
	\left(\partial_{\lambda_\alpha}{\bf B}^{}_{\bm\lambda}\right)
	{\bf G}^{}_{\bm\lambda}
\end{equation}.
The second derivative of ${\bf G}^{}({\bm\lambda})$ follows by induction as
\begin{align}
	\partial_{\lambda_\alpha}\partial_{\lambda_\beta}{\bf G}^{}_{\bm\lambda}
	&=
	\left(\partial_{\lambda_\beta}{\bf G}^{}_{\bm\lambda}\right)
	\left(\partial_{\lambda_\alpha}{\bf B}^{}_{\bm\lambda}\right)
	{\bf G}^{}_{\bm\lambda}
	+
	{\bf G}^{}_{\bm\lambda}
	\left(\partial_{\lambda_\beta}\partial_{\lambda_\alpha}{\bf B}^{}_{\bm\lambda}\right)
	{\bf G}^{}_{\bm\lambda}
	+
	{\bf G}^{}_{\bm\lambda}
	\left(\partial_{\lambda_\alpha}{\bf B}^{}_{\bm\lambda}\right)
	\left(\partial_{\lambda_\beta}{\bf G}^{}_{\bm\lambda}\right)\\
	&=
	{\bf G}^{}_{\bm\lambda}
	\left(\partial_{\lambda_\beta}{\bf B}^{}_{\bm\lambda}\right)
	{\bf G}^{}_{\bm\lambda}
	\left(\partial_{\lambda_\alpha}{\bf B}^{}_{\bm\lambda}\right)
	{\bf G}^{}_{\bm\lambda}
	+
	{\bf G}^{}_{\bm\lambda}
	\left(\partial_{\lambda_\beta}\partial_{\lambda_\alpha}{\bf B}^{}_{\bm\lambda}\right)
	{\bf G}^{}_{\bm\lambda}
	+
	{\bf G}^{}_{\bm\lambda}
	\left(\partial_{\lambda_\alpha}{\bf B}^{}_{\bm\lambda}\right)
	{\bf G}^{}_{\bm\lambda}
	\left(\partial_{\lambda_\beta}{\bf B}^{}_{\bm\lambda}\right)
	{\bf G}^{}_{\bm\lambda}.
\end{align}

Using the above identity ${\bf B}^{u}{\bf G}^{}_{\bm\lambda}|_{\bm\lambda=\bf0}={\bf 1}$ we find
$
\left[\langle{\bf x}\rangle\right]_\alpha
 = {\bf 1}\left(\partial_{\lambda_\alpha}
{\bf B}^{}_{\bm\lambda}\right)
{\bf G}^{}_{\bm\lambda}
{\bf P}(0)\Big|_{{\bm\lambda}=\bf0}$ and
\begin{equation}
	\left[\langle{\bf x}\otimes{\bf x}\rangle\right]_{\alpha\beta}
	=
	{\bf 1}
	\left[
	\left(\partial_{\lambda_\beta}{\bf B}^{}_{\bm\lambda}\right)
	{\bf G}^{}_{\bm\lambda}
	\left(\partial_{\lambda_\alpha}{\bf B}^{}_{\bm\lambda}\right)
	+
	\left(\partial_{\lambda_\beta}\partial_{\lambda_\alpha}{\bf B}^{}_{\bm\lambda}\right)
	+
	\left(\partial_{\lambda_\alpha}{\bf B}^{}_{\bm\lambda}\right)
	{\bf G}^{}_{\bm\lambda}
	\left(\partial_{\lambda_\beta}{\bf B}^{}_{\bm\lambda}\right)
	\right]
	{\bf G}^{}_{\bm\lambda}
	{\bf P}(0)
	\Big|_{{\bm\lambda}=\bf0}.
\end{equation}
We now evaluate the vector and matrix-valued first and second derivatives of the components of ${\bf B}^{}_{\bm\lambda}$, writing
\begin{equation}
	\left[\partial_{\bm\lambda}
	{\bf B}^{}_{\bm\lambda}
	|_{{\bm\lambda}=\bf0}
	\right]_{qp}
	=
	\sum_{l\in\mathcal{C}_{qp}}
	\frac{{\rm k}_{l}}{{\rm k}^{tot}_p}
	%\left[{\bf d}_{l}\right]_\alpha
	{\bf d}_{l}
	\equiv
	\left[\overline{{\bf B}\otimes{\bf d}}\right]_{qp},
	\quad
	\left[
	\partial_{\bm\lambda}\otimes\partial_{\bm\lambda}
	{\bf B}^{}({\bm\lambda})
	|_{{\bm\lambda}=\bf0}
	\right]_{qp}
	=
	\sum_{l\in\mathcal{C}_{qp}}
	\frac{{\rm k}_{l}}{{\rm k}^{tot}_p}
	{\bf d}_l\otimes{\bf d}_l
	%\left[{\bf d}_{l}\right]_\alpha
	%\left[{\bf d}_{l}\right]_\beta
	\equiv
	\left[\overline{{\bf B}\otimes{\bf d}\otimes{\bf d}}\right]_{qp}.
\end{equation}

\section{Transport coefficients from trajectories of variable duration}
As discussed in the main text and below, in the well sampled limit the matrix ${\bf K}^{tot}-{\bf K}^{}$ has a spectral gap, with one eigenvalue $\nu_0>0$ much smaller than all the others. The corresponding right eigenvector is the quasistationary distribution\cite{lebris2012} ${\bm\pi}^{QSD}$, which as $\nu_0\to0$ becomes the Boltzmann distribution $\hat{\bm\pi}$ in equilibrium. The QSD is a natural choice of initial distribution, as it is the limiting distribution in the known state space conditional on not absorbing. As a result, setting
${\bf P}(0)=\hat{\bm\pi}^{QSD}$ effectively eliminates initial transients in the extraction of transport coefficients.
With this choice of initial condition the residence time distribution $\dot{\rm P}(\tau)={\bf 1}\dot{\bf P}(\tau)$ is a single exponential with decay constant $\nu_0$, i.e. $\dot{\rm P}(\tau)=\nu_0\exp(-\nu_0\tau)$.
Consider an average $\langle\dots\rangle_\tau$ over the set of all trajectories $\mathcal{P}_\tau$ with $\tau_{res}\in[\tau,\tau+{\rm d}\tau]$.
It is clear that $\langle\dots\rangle=\int_0^\infty\dot{\rm P}(\tau)\langle\dots\rangle_\tau{\rm d}\tau$.
The drift and diffusion estimators ${\bm\mu}_\tau$ and ${\bf D}_\tau$ over such trajectories are defined through
\begin{equation}
	\langle{\bf x}\rangle_\tau = \sum_{\xi\in\mathcal{P}_\tau}{\bf x}_\xi P_\xi \equiv \tau{\bm\mu}_\tau
	,\quad
	\langle{\bf x}\otimes{\bf x}\rangle_\tau =
	\sum_{\xi\in\mathcal{P}_\tau}{\bf x}_\xi\otimes{\bf x}_\xi P_\xi = 2\tau{\bf D}_\tau + \tau^2{\bm\mu}_\tau\otimes{\bm\mu}_\tau.
\end{equation}

With increasing trajectory length, $\tau\to\infty$, the drift and diffusion coefficients are only well defined if the estimators tend to a constant value, i.e.
$\lim_{\tau\to\infty}{\bm\mu}_\tau = {\bm\mu}$ and $\lim_{\tau\to\infty}{\bf D}_\tau = {\bf D}$.

In the well sampled limit where ${\bf K}^{tot}-{\bf K}^{}$ has a spectral gap, averages over the set of all possible paths $\mathcal{P}=\cup_\tau\mathcal{P}_\tau$ will be dominated by the large $\tau$ regime, where ${\bm\mu}_\tau$ and ${\bf D}_\tau$ are constant.
Furthermore, due the preparation of the initial conditions in the QSD (${\bf P}(0)=\hat{\bm\pi}^{QSD}$) the exit time distribution is the single exponential decay
$\dot{\rm P}(\tau)=\nu_0\exp(-\nu_0\tau)$, meaning $\langle\tau\rangle=\nu_0^{-1}$ and $\langle\tau^2\rangle=2\nu_0^{-2}=2\langle\tau\rangle^2$.\\

We can therefore extract ${\bm\mu},{\bf D}$ through
\begin{align}
	\langle{\bf x}\rangle &=
	%\lim_{\langle\tau\rangle\to\infty}
	\int_0^\infty\dot{\rm P}(\tau)\tau{\bm\mu}_\tau{\rm d}\tau
	\equiv\langle\tau\rangle{\bm\mu}
	,\\
	%\lim_{\langle\tau\rangle\to\infty}
	\langle{\bf x}\otimes{\bf x}\rangle &=
	%\lim_{\langle\tau\rangle\to\infty}
	\int_0^\infty\dot{\rm P}(\tau)\left(2\tau{\bf D}_\tau+\tau^2{\bm\mu}_\tau\otimes{\bm\mu}_\tau\right){\rm d}\tau
	\equiv
	2\langle\tau\rangle{\bf D} + \langle\tau^2\rangle{\bm\mu}\otimes{\bm\mu}
	=
	2\langle\tau\rangle{\bf D} + 2\langle\tau\rangle^2{\bm\mu}\otimes{\bm\mu}
	.
\end{align}
Exploiting the chain rule and the identity $Z({\bf0})=1$, the drift and diffusion constants can thus be succinctly written
\begin{align}
	%Z({\bm\lambda}) &= {\bf B}^{u}{\bf G}^{}({\bm\lambda})\hat{\bm\pi}^{QSD}\\
	{\bm\mu} &= \lim_{\langle\tau\rangle\to\infty} \frac{\langle{\bf x}\rangle}{\langle\tau\rangle}
	= -\lim_{{\bf k}^{u}\to{\bf0}} \langle\tau\rangle^{-1}\partial_{\bm\lambda}Z({\bm\lambda})%\Big|_{{\bm\lambda=\bf0},{\bf P}(0)=\hat{\bm\pi}^{QSD}}
	= \lim_{{\bf k}^{u}\to{\bf0}}\langle\tau\rangle^{-1}\partial_{\bm\lambda}\left[1/Z({\bm\lambda})\right]\Big|_{{\bm\lambda=\bf0},{\bf P}(0)=\hat{\bm\pi}^{QSD}},
	\\
	{\bf D} &=
	\lim_{\langle\tau\rangle\to\infty}
	\left(
	\frac{\langle{\bf x}\otimes{\bf x}\rangle}{2\langle\tau\rangle}
	-
	\langle\tau\rangle
	{\bm\mu}\otimes{\bm\mu}
	\right)
	=
	\frac{1}{2}
	\lim_{{\bf k}^{u}\to{\bf0}}
	\langle\tau\rangle^{-1}\partial_{\bm\lambda}\otimes\partial_{\bm\lambda}\left[1/Z({\bm\lambda})\right]\Big|_{{\bm\lambda=\bf0},{\bf P}(0)=\hat{\bm\pi}^{QSD}}.
\end{align}
as quoted in the main text. We thus see that when trajectories are of variable length, the appropriate expression to extract the diffusivity in the possible presence of drift is not simply the variance in the total displacement, but must account for the properties of the distribution of trajectory durations.
%In the present case, for a well sampled model the trajectory duration distribution is an exponential decay, allowing the analytical replacement $\langle\tau^2_{res}\rangle=2\langle\tau_{res}\rangle^2$.
\section{Explicit expressions and limits for transport coefficients}
To extract transport coefficients we analytically evolve the system from the QSD, i.e. ${\bf P}(0)=\hat{\bm\pi}^{QSD}$, where ${\bm\pi}^{QSD}$ is the right eigenvector ${\bf v}_0$ of the matrix ${\bf K}^{tot}-{\bf K}^{}$ with eigenvalue $\nu_0>0$ much smaller than all the others. The choice of notation is motivated by the limiting form of the QSD in equilibrium, namely the Boltzmann distribtion: in the complete sampling limit $\nu_0\to0$
${\bm\pi}^{QSD}\to\hat{\bm\pi}$, in equilibrium, where the curcumflex implies $L_1$ normalization. We also label the corresponding left eigenvector as
${\bf w}_0={\bf 1}^{QSD}$ which has the $\nu_0\to0$ limit ${\bf 1}^{QSD}\to{\bf1}$.

The probability density in the known state space becomes the single exponential decay
\begin{equation}
	\dot{\rm P}(t) = {\bf 1}\dot{\bf P}(t) = \sum_l\nu_l\exp(-\nu_l\tau)\left({\bf 1}{\bf v}_l\right)\left({\bf w}_l{\bm\pi}^{QSD}\right)=\nu_0\exp(-\nu_0t).
\end{equation}
The expected residence time is thus $\langle\tau_{res}\rangle=\nu^{-1}_0$, as disucussed above.

Greens function matrix ${\bf G}^{}$ acting on $\hat{\bm\pi}^{QSD}$ gives simply
\begin{equation}
	{\bf G}^{}\hat{\bm\pi}^{QSD} = \langle\tau_{res}\rangle{\bf K}^{tot}\hat{\bm\pi}^{QSD}
\end{equation}
by the orthonormality relation ${\bf w}_l{\bf v}_m = \delta_{lm}$.

Expanding $\langle{\bf x}\rangle$ under the QSD initial condition we find the limiting expression
\begin{equation}
	\langle{\bf x}\rangle =
	\langle\tau_{res}\rangle
	{\bf 1}
	\overline{{\bf B}\otimes{\bf d}}
	\,{\bf K}^{tot}
	\hat{\bm\pi}^{QSD}
	,\quad\Rightarrow\quad
	{\bm\mu}({\bf k}^{u})=
	{\bf 1}\overline{{\bf B}\otimes{\bf d}}\,{\bf K}^{tot}
	\hat{\bm\pi}^{QSD}
	\to
	{\bf 1}\overline{{\bf B}\otimes{\bf d}}\,{\bf K}^{tot}
	\hat{\bm\pi}={\bm\mu}
\end{equation}
Whilst in equilibrium $\bm\mu={\bf0}$, we retain the general case to treat drift problems.
The second moment reads
\begin{equation}
	\langle{\bf x}\otimes{\bf x}\rangle = {\bf M}+{\bf M}
	,\quad
	{\bf M}
	=
	\langle\tau_{res}\rangle
	{\bf 1}
	\left[
	\frac{1}{2}\overline{{\bf B}\otimes{\bf d}\otimes{\bf d}}
	+
	\overline{{\bf B}\otimes{\bf d}}
	{\bf G}^{}
	\overline{{\bf B}\otimes{\bf d}}
	\right]
	{\bf K}^{tot}
	\hat{\bm\pi}^{QSD}.
\end{equation}
It is then possible to define diffusion coefficients through the above relation
\begin{align}
	{\bf D}({\bf k}^{u})
	&=
	\frac{1}{2\langle\tau_{res}\rangle}
	\left[
	\langle{\bf x}\otimes{\bf x}\rangle - 2\langle{\bf x}\rangle\otimes\langle{\bf x}\rangle
	\right]
	=
	\frac{1}{2}
	\left[
	\widetilde{\bf D}({\bf k}^{u})+\widetilde{\bf D}^\top({\bf k}^{u})
	\right]
	\\
	\widetilde{\bf D}({\bf k}^{u}) &=
	{\bf 1}
	\left[
	\frac{1}{2}\overline{{\bf B}\otimes{\bf d}\otimes{\bf d}}
	+
	\overline{{\bf B}\otimes{\bf d}}
	\left({\bf G}^{}-
	\langle\tau_{res}\rangle{\bf K}^{tot}\hat{\bm\pi}^{QSD}
	\otimes{\bf 1}\right)
	\overline{{\bf B}\otimes{\bf d}}
	\right]
	{\bf K}^{tot}
	\hat{\bm\pi}^{QSD}
\end{align}
However, as $\nu_0\to0$ the Greens function matrix ${\bf G}^{}$ is singular. We make a decomposition into a nonsingular and singular part as
\begin{equation}
	{\bf G}^{} = {\bf K}^{tot}\left[{\bf K}^{tot}-{\bf K}^{}\right]^{-1} = {\bf G}_+^{} +
	\langle\tau_{res}\rangle
	{\bf K}^{tot}{\bm\pi}^{QSD}\otimes{\bf 1}^{QSD},
\end{equation}
where ${\bf G}^{}_+$ is thus the \textit{pseudoinverse} of $\left[\mathbb{I}-{\bf B}^{}\right] = \left[{\bf K}^{tot}-{\bf K}^{}\right]\left[{\bf K}^{tot}\right]^{-1}$ in the limit $\nu_0\to0$.
This gives the well defined limit of
\begin{equation}
	\lim_{{\bf k}^{u}\to{\bf0}} \widetilde{\bf D}({\bf k}^{u})
	=
	{\bf 1}
	\left[
	\frac{1}{2}
	\overline{{\bf B}\otimes{\bf d}\otimes{\bf d}}
	+
	\overline{{\bf B}\otimes{\bf d}}\,{\bf G}^{}_+\overline{{\bf B}\otimes{\bf d}}
	\right]{\bf K}^{tot}\hat{\bm\pi}=
	\widetilde{\bf D},
	\quad
	{\bf D} =
	\frac{1}{2}
	\left[
	\widetilde{\bf D}+\widetilde{\bf D}^\top
	\right].
\end{equation}
Recognizing that ${\bf K}^{tot}{\bf G}^{}_+$ is precisely the pseudoinverse ${\bf W}^+$ of the singular matrix
$\lim_{\nu_0\to0}\left[{\bf K}^{tot}-{\bf K}^{}\right]$, where the choice of notation follows Trinkle\cite{Trinkle2018}, we can write the limiting
form for $\widetilde{\bf D}$ as
\begin{equation}
	\widetilde{\bf D}
	=
	\frac{1}{2}
	{\bf 1}
	\overline{{\bf K}^{}\otimes{\bf d}\otimes{\bf d}}\hat{\bm\pi}
	+
	{\bf 1}
	\overline{{\bf K}^{}\otimes{\bf d}}\,{\bf W}^+\overline{{\bf K}^{}\otimes{\bf d}}
	\hat{\bm\pi},
\end{equation}
where
\begin{equation}
	\left[\overline{{\bf K}^{}\otimes{\bf d}}\right]_{qp} \equiv
	\sum_{l\in\mathcal{C}_{qp}}
	{\rm k}_{l} {\bf d}_{l}
	,\quad
	\left[\overline{{\bf K}^{}\otimes{\bf d}\otimes{\bf d}}\right]_{qp} \equiv
	\sum_{l\in\mathcal{C}_{qp}}
	{\rm k}_{l} {\bf d}_{l}\otimes{\bf d}_{l}.
\end{equation}
In this notation, ${\bm\mu}={\bf1}\overline{{\bf K}^{}\otimes{\bf d}}{\bm\pi}$. Elementary algebraic manipulations show that ${\rm Tr}{\bf D}$ gives exactly the
form for the self diffusion constant (diagonal terms in the Onsager matrix) as derived recently by Trinkle\cite{Trinkle2018}$^,$\footnote{See equation S15 of the supplementary material for the most direct comparison}.
We also refer the reader to many other works on diffusion in periodic media\cite{landman1979stochastic}.

\section{Symmetries under detailed balance}
With a steady state Boltzmann distribution $\lim_{\nu_0\to0}\left[{\bf K}^{tot}-{\bf K}^{}\right]{\bm\pi}=\bf0$, and a Boltzmann matrix
$\left[{\bm\Pi}\right]_{pq} = \delta_{pq}\pi_q$, the detailed balance condition reads
\begin{equation}
	{\bf K}^{}{\bm\Pi} = \left[{\bf K}^{}{\bm\Pi}\right]^\top = {\bm\Pi}\left[{\bf K}^{}\right]^\top
	,\quad\Rightarrow\quad
	{\bm\Pi}^{-1}{\bf K}^{} =\left[{\bf K}^{}\right]^\top{\bm\Pi}^{-1}
	,\quad\Rightarrow\quad
	{\bf W}^{+}{\bm\Pi} = {\bm\Pi}\left[{\bf W}^{+}\right]^\top.
	%\delta{\bf K}^{}{\bm\Pi} = {\bm\Pi}\left[\delta{\bf K}^{}\right]^\top.
\end{equation}
We define the transpose of $\overline{{\bf K}^{}\otimes{\bf d}}$ and $\overline{{\bf K}^{}\otimes{\bf d}\otimes{\bf d}}$ as
\begin{equation}
	\left[\overline{{\bf K}^{}\otimes{\bf d}}^\top\right]_{qp}
	\equiv
	\sum_{l\in\mathcal{C}_{pq}}
	{\rm k}_{l} {\bf d}_{l}
	,\quad
	\left[\overline{{\bf K}^{}\otimes{\bf d}\otimes{\bf d}}^\top\right]_{qp}
	\equiv
	\sum_{l\in\mathcal{C}_{pq}}
	{\rm k}_{l} {\bf d}_{l}\otimes{\bf d}_l.
\end{equation}
%with analagous expressions for $\overline{\delta{\bf K}^{}\otimes{\bf d}}$ and $\overline{\delta{\bf K}^{}\otimes{\bf d}\otimes{\bf d}}$.
%
As every jump vector ${\bf d}_l,l\in\delta\mathcal{C}_{pq}$ will have a one-to-one correspondence with a vector ${\bf d}_{l'}=-{\bf d}_l$ for some $l'\in\delta\mathcal{C}_{qp}$, the detailed balance symmetries imply in particular that
\begin{equation}
	\overline{{\bf K}^{}\otimes{\bf d}}{\bm\Pi} = -{\bm\Pi}\overline{{\bf K}^{}\otimes{\bf d}}^\top
	,\quad
	\overline{{\bf K}^{}\otimes{\bf d}\otimes{\bf d}}{\bm\Pi} = {\bm\Pi}\overline{{\bf K}^{}\otimes{\bf d}\otimes{\bf d}}^\top,
	\label{db2}.
\end{equation}
%with again analagous expressions for $\overline{\delta{\bf K}^{}\otimes{\bf d}}$ and $\overline{\delta{\bf K}^{}\otimes{\bf d}\otimes{\bf d}}$. \\

These relations imply that the drift vector is always zero in equilibrium- using ${\bm\pi}={\bm\Pi}{\bf1}^\top$ and the fact that
${\bf 1}{\bf M}{\bf 1}^\top = {\bf 1}{\bf M}^\top{\bf 1}^\top$ for any matrix ${\bf M}$ we have
\begin{equation}
	{\bm\mu}
	=
	{\bf 1}\overline{{\bf K}^{}\otimes{\bf d}}{\bm\Pi}{\bf 1}^\top={\bf 1}\left[\overline{{\bf K}^{}\otimes{\bf d}}{\bm\Pi}\right]^\top{\bf 1}^\top
	=-{\bf 1}\overline{{\bf K}^{}\otimes{\bf d}}{\bm\Pi}{\bf 1}^\top=-{\bm\mu}
	,\quad\Rightarrow
	{\bm\mu}={\bf 0}.
	%,\quad\Rightarrow
	%\delta{\bm\mu}={\bf 0},
\end{equation}
%where the last equality is due to the detailed balance symmetry of $\overline{\delta{\bf K}^{}\otimes{\bf d}}$.\\
We now split $\widetilde{\bf D}$ into uncorrelated and correlated terms as
\begin{equation}
	\widetilde{\bf D} = \widetilde{\bf D}_{uc} + \widetilde{\bf D}_c
	,\quad
	\widetilde{\bf D}_{uc} = \frac{1}{2}{\bf 1}\overline{{\bf K}^{}\otimes{\bf d}\otimes{\bf d}}{\bm\Pi}{\bf1}^\top
	,\quad
	\widetilde{\bf D}_{c}
	=
	{\bf 1}
	\overline{{\bf K}^{}\otimes{\bf d}}\,{\bf W}^+\overline{{\bf K}^{}\otimes{\bf d}}
	{\bm\Pi}{\bf1}^\top.
\end{equation}
As all rates are positive, it is clear that ${\rm Tr}\widetilde{\bf D}_{uc} >0$. For the correlated contribution we first use the symmetry of ${\bf W}^+$
to write
\begin{equation}
	{\bm\Pi}^{-1}{\bf W}^+ = \left[{\bm\Pi}^{-1}{\bf W}^+\right]^\top \equiv \widetilde{\bf W}^\top\widetilde{\bf W}
\end{equation}
where we introduce $\widetilde{\bf W}$ to write the symmetric matrix ${\bm\Pi}^{-1}{\bf W}^+$ as $\widetilde{\bf W}^\top\widetilde{\bf W}$.
We then use detailed balance symmetries to rewrite the correlated contribution as
\begin{equation}
	%\widetilde{\bf D}_c =
	%-{\bf 1}{\bm\Pi}
	%\overline{{\bf K}^{}\otimes{\bf d}}^\top{\bm\Pi}^{-1}{\bf W}^+\overline{{\bf K}^{}\otimes{\bf d}}
	%{\bm\Pi}{\bf1}^\top
	%,\quad
	\left[\widetilde{\bf D}_c\right]_{\alpha\beta} = -\sum_p {\rm A}_{p\alpha}{\rm A}_{p\beta}
	,\quad
	{\bf A} = \widetilde{\bf W}\overline{{\bf K}^{}\otimes{\bf d}}{\bm\Pi}{\bf1}^\top
	= -\widetilde{\bf W}{\bm\Pi}\overline{{\bf K}^{}\otimes{\bf d}}^\top{\bf1}^\top
	,\quad\Rightarrow
	\left[\widetilde{\bf D}_{c}\right]_{\alpha\alpha}
	=
	-\sum_p {\rm A}^2_{p\alpha} < 0
	,\quad\Rightarrow
	{\rm Tr} \widetilde{\bf D}_{c} < 0,
\end{equation}
which shows that the correlated contribution always reduces the diffusivity for a system obeying detailed balance.

%\subsection{Analytic sensitivity- don't think we need this now}
%Using ${\bf W}^+ = {\bm\Pi}\widetilde{\bf W}^\top\widetilde{\bf W}$ we find that
%\begin{equation}
%	{\bf W}^+\overline{{\bf K}^{}\otimes{\bf d}}{\bm\Pi}{\bf1}^\top
%	=
%	{\bm\Pi}\widetilde{\bf W}^\top{\bf A}
%	\quad,
%	%
%	{\bf1}\overline{{\bf K}^{}\otimes{\bf d}}{\bf W}^+
%	=
%	-{\bf A}^\top\widetilde{\bf W}
%\end{equation}
%Defining $\delta{\bf A}\equiv \widetilde{\bf W}\overline{\delta{\bf K}^{}\otimes{\bf d}}{\bm\Pi}{\bf1}^\top$,
%the diffusion tensor ${\bf D}=\frac{1}{2}\left[\widetilde{\bf D}+\widetilde{\bf D}^\top\right]$ undergoes pertubations given by
%\begin{align}
%	\delta\widetilde{\bf D}
%	&\to
%	\frac{1}{2}
%	{\bf 1}
%	\overline{\delta{\bf K}^{}\otimes{\bf d}\otimes{\bf d}}{\bm\Pi}{\bf 1}^\top
%	-
%	2{\bf A}^\top\delta{\bf A}
%	-{\bf A}^\top\widetilde{\bf W}\delta{\bf K}{\bm\Pi}\widetilde{\bf W}^\top{\bf A}
%	\label{dD}
%\end{align}
%where we use $\delta{\bf W}^+ = {\bf W}^+\left(\delta{\bf K}^{}-\delta{\bf K}^{tot}\right){\bf W}^+$.
%
\section{Sensitivity of the diffusivity tensor to additional sampling data}
Under additional sampling, the unknown rates ${\bf k}^{u}$ will decrease, additional transition rates will be found, and new states may be discovered.
We first focus on the case where no new state is discovered, only additional rates, and assume the QSD $\hat{\bm\pi}^{QSD}\simeq\hat{\bm\pi}$ is unchanged.
The rate matrix pertubations can be grouped into additional sets of connections $\mathcal{C}_{pq}\to\mathcal{C}_{pq}+\delta\mathcal{C}_{pq}$ and changes in
the unknown rates $\delta{\rm k}^{u}$, satisfying
\begin{equation}
	\left[\delta{\bf K}^{}\right]_{pq} = \sum_{l\in\delta\mathcal{C}_{pq}}{\rm k}_l>0
	,\quad
	\left[\delta{\bf k}^{u}\right]_p \leq -\left[{\bf 1}\delta{\bf K}^{}\right]_p
	,
	\left[\delta{\bf k}^{u}\right]_p \leq -\left[{\bf k}^{u}\right]_p
	,\quad
	\Rightarrow
	\quad
	\left[\delta{\bf K}^{tot}\right]_{pp} \leq {0},
\end{equation}
which assumes that the unknown rate estimates ${\bf k}^{u}$ are accurate and therefore an upper bound, meaning that the total additional escape rates
${\bf 1}\delta{\bf K}^{}$ from each state cannot exceed the negative change in unknown rates $-\delta{\bf k}^{u}$.
%
%Taking the limit ${\bf 1}\delta{\bf K}^{}=-\delta{\bf k}^u={\bf k}^u$,
%\subsection{Monte Carlo procedure for only additional rates}
To generate candidate $\delta{\bf K}$ we first define the symmetric matrix $\delta\widetilde{\bf K}$ which satisfies
\begin{equation}
	\delta\widetilde{\bf K}=\delta\widetilde{\bf K}^\top
	,\quad
	\delta\widetilde{\bf K} = \delta{\bf K}{\bm\Pi}
	,\quad
	\left[{\bf1}\delta\widetilde{\bf K}\right]_p \leq \left[{\bm\Pi}{\bf k}^u\right]_p
\end{equation}
The Monte Carlo procedure discussed in the main text created candidate $\delta{\bf K}$ in the following procedure-
\begin{enumerate}
	\item $\mathcal{L}$ = randomly shuffled list of pairs $p,q$, where $p\leq q$
	\item Iterate through $\mathcal{L}$, filling $\delta\widetilde{\rm K}_{pq}=\delta\widetilde{\rm K}_{qp}$ with a
	random number drawn uniformly from $[0,\min({\rm k}^u_q\pi_q-\sum_p\widetilde{\rm K}_{pq},{\rm k}^u_p\pi_p-\sum_q\widetilde{\rm K}_{qp}))$, to ensure the above inequality is satisfied
	\item Pick a random (possibly zero) lattice vector for each new transition add this to the difference in
	intercell positions to form ${\bf d}_{pq}=-{\bf d}_{qp}$ for each new transition
	\item Add $\delta{\bf K}=\delta\widetilde{\bf K}{\bm\Pi}^{-1}$ to existing ${\bf K}$, remove ${\bf1}\delta{\bf K}$ from ${\bf k}^u$ then calculate new ${\bf D}+\delta{\bf D}$
	\item Repeat until convergence
\end{enumerate}
We found around 300 samples per temperature were adequate, though in our simple Python implementation each iteration took around 20$\mu$s on a single
Intel i5-8250U CPU core, with well converged sampling for the entire temperature range of interest taking around 15-20 seconds.

\end{document}